\documentclass{article}

\usepackage{arxiv}

\usepackage{tgtermes}
\usepackage{tgheros}
\usepackage{tgcursor}
\usepackage[T1]{fontenc}
\usepackage{iftex}
\ifPDFTeX
  \usepackage[utf8]{inputenc}
  \DeclareUnicodeCharacter{2013}{--}
  \DeclareUnicodeCharacter{2014}{---}
  \DeclareUnicodeCharacter{2018}{`}
  \DeclareUnicodeCharacter{2019}{'}
  \DeclareUnicodeCharacter{201C}{``}
  \DeclareUnicodeCharacter{201D}{''}
  \DeclareUnicodeCharacter{2026}{\ldots}
  \DeclareUnicodeCharacter{00D7}{\ensuremath{\times}}
  \DeclareUnicodeCharacter{00B1}{\ensuremath{\pm}}
  \DeclareUnicodeCharacter{2192}{\ensuremath{\to}}
  \DeclareUnicodeCharacter{2190}{\ensuremath{\leftarrow}}
  \DeclareUnicodeCharacter{2264}{\ensuremath{\leq}}
  \DeclareUnicodeCharacter{2265}{\ensuremath{\geq}}
  \DeclareUnicodeCharacter{03B1}{\ensuremath{\alpha}}
  \DeclareUnicodeCharacter{03B2}{\ensuremath{\beta}}
  \DeclareUnicodeCharacter{03B3}{\ensuremath{\gamma}}
  \DeclareUnicodeCharacter{03B4}{\ensuremath{\delta}}
  \DeclareUnicodeCharacter{03B5}{\ensuremath{\varepsilon}}
  \DeclareUnicodeCharacter{03B8}{\ensuremath{\theta}}
  \DeclareUnicodeCharacter{03BB}{\ensuremath{\lambda}}
  \DeclareUnicodeCharacter{03BC}{\ensuremath{\mu}}
  \DeclareUnicodeCharacter{03C0}{\ensuremath{\pi}}
  \DeclareUnicodeCharacter{03C3}{\ensuremath{\sigma}}
\fi

\usepackage{amsmath}
\usepackage{amssymb}
\usepackage{amsthm}
\usepackage{mathtools}
\usepackage{amsfonts}

\usepackage{microtype}
\usepackage{enumitem}

\usepackage{graphicx}
\usepackage{booktabs}
\usepackage{array}
\usepackage{tabularx}
\usepackage{multirow}
\usepackage{caption}
\usepackage{subcaption}
\usepackage{xcolor}
\usepackage{listings}
\usepackage{tikz}
\usetikzlibrary{positioning,arrows.meta,shapes.geometric,shapes.symbols,shapes.misc,fit,backgrounds,calc,decorations.pathreplacing,patterns}

\usepackage{pdfpages}

\usepackage[breakable,skins]{tcolorbox}
\tcbset{
  ideabox/.style={
    enhanced, breakable, colback=blue!2, colframe=blue!35!black,
    boxrule=0.4pt, arc=2pt, left=4pt, right=4pt, top=3pt, bottom=3pt,
    fonttitle=\bfseries\small, coltitle=blue!35!black,
    attach boxed title to top left={xshift=4mm, yshift=-2.4mm},
    boxed title style={colback=white, colframe=blue!35!black, boxrule=0.4pt,
                       arc=2pt, left=3pt, right=3pt, top=1pt, bottom=1pt},
    title=\strut,
  },
}

\usepackage{url}
\usepackage{hyperref}
\hypersetup{
  colorlinks=true,
  linkcolor=blue!50!black,
  citecolor=blue!50!black,
  urlcolor=blue!50!black,
}

\usepackage{authblk}

\newcommand{\parness}{\textsf{PARNESS}}
\newcommand{\code}[1]{\texttt{\small #1}}

\title{\textsf{PARNESS}: A Paper Harness for End-to-End Automated\\
  Scientific Research with Dynamic Workflows, Full-Text Indexing,\\
  and Cross-Run Knowledge Accumulation}

\author[1]{Yuchen Wang\thanks{First author and corresponding author.
  \texttt{wangyuchen21@buaa.edu.cn}}}
\author[1]{Zhongzhi Luan\thanks{Corresponding author.
  \texttt{rick710055@263.net}}}
\affil[1]{Sino-German Joint Software Institute, Beihang University, Beijing, China}

\renewcommand{\shorttitle}{\textsf{parness}: A Paper Harness for Automated Research}

\hypersetup{
  pdftitle={parness: A Paper Harness for End-to-End Automated Scientific Research},
  pdfsubject={cs.AI, cs.SE},
  pdfauthor={Yuchen Wang, Zhongzhi Luan},
  pdfkeywords={parness, paper harness, automated research, LLM agents, DAG, multi-agent, knowledge graph, pipeline DSL, research life-cycle},
}

\date{May 2026}

\begin{document}
\maketitle

\begin{abstract}
\noindent
Recent autonomous research systems --- AI-Scientist v1/v2~\cite{aisciv1,aisciv2},
PaperOrchestra~\cite{paperorchestra}, OpenAGS, Karpathy's
\emph{autoresearch}, AutoSOTA~\cite{autosota}, Tongyi DeepResearch~\cite{deepresearch},
InternAgent~\cite{internagent} and ResearchAgent~\cite{researchagent} --- show
that LLM agents can ideate, run experiments and write papers, but each fixes
a particular control-flow shape (linear pipeline, hard-coded state machine,
single-agent loop, or skill-pack with a fixed five-agent recipe) at the
framework level. We argue that this rigidity has \emph{five} structural
roots: \textbf{(1)} research workflows are dynamic and discipline-specific
(lab experiments, surveys, simulations, theory, ablations all loop
differently and discuss differently); \textbf{(2)} ideation is bounded by
LLM context~\cite{lostmiddle} and cross-domain ideation depends on
accumulated knowledge a single context cannot hold;
\textbf{(3)} relying on paper summaries alone misses what is in the body,
yet legal full-text access is uneven --- so the \emph{cumulative} corpus,
not the \emph{instant} corpus, must do the work; \textbf{(4)} a paper's
open-source repository (typically on GitHub) is often the only complete
specification of its experimental scheme, and the paper$\leftrightarrow$code
correspondence is a first-class research artefact that current systems
neglect; \textbf{(5)} no existing tool persists cross-run knowledge in a
form that can be retrieved into a finite LLM context.

We present \parness{}\footnote{\textsf{PARNESS} stands for \emph{paper
harness}: a thin runtime that binds heterogeneous research components
into a single declarative pipeline. Source code and pipeline
configurations: \url{https://github.com/gtrhythm/PARNESS}.}, an
open-source framework built on
four design moves that map directly onto these five roots. \emph{(i)} A
thin DAG kernel with a four-field Agent contract decouples scheduling
from domain semantics, so any discipline's loop and discussion mode is
expressible as user-editable YAML rather than orchestrator code.
\emph{(ii)} A full-text PDF-parsing and literature-library subsystem
indexes paper bodies, figures and tables as typed objects, with
graceful abstract-only fall-back so the system's reading capability
grows monotonically with use. \emph{(iii)} A knowledge-graph index
over papers, \emph{ideas}, \emph{experiments} and \emph{code
repositories}, with scenario-typed retrieval (similar / contradictory
/ cross-domain / counter-intuitive), surfaces a focused slice of the
cumulative corpus into each LLM call. \emph{(iv)} A deliberately small
extension surface lets any modern coding agent (Claude Code, Cursor,
Copilot, OpenCode, Kilo Code) add, remove or replace any module
without a custom plug-in. The reference implementation runs the full
research life-cycle end-to-end on real document collections; to our
knowledge it is the first open-source system that combines declarative
pipeline composition, full-PDF + code-repository indexing, and
persistent cross-run knowledge in a single codebase.
\end{abstract}


\section{Introduction}
\label{sec:intro}

\begin{figure}[t]
\centering
\includegraphics[width=0.95\linewidth]{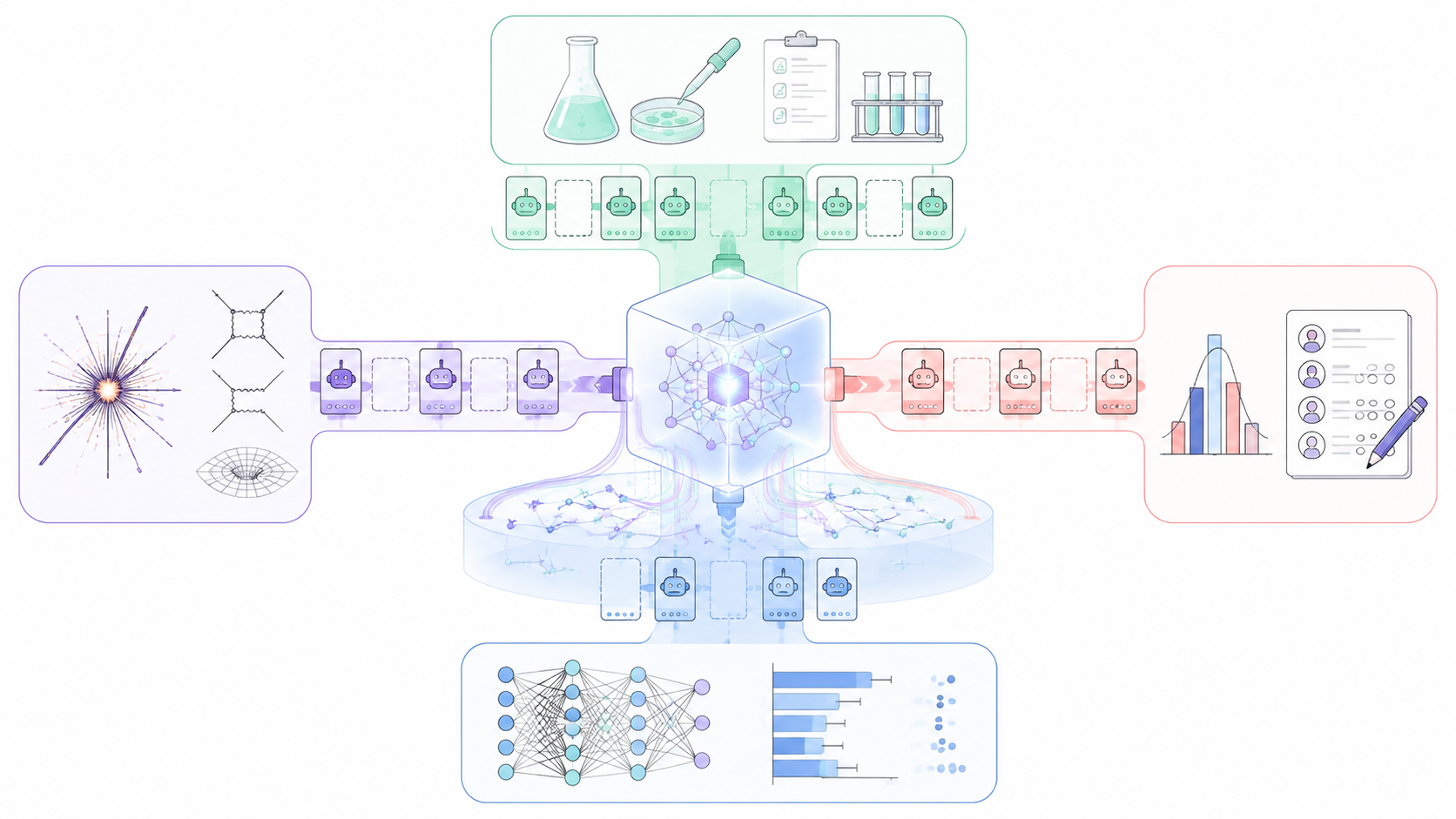}
\caption{\parness{} as a paper harness. A single thin DAG kernel
(centre) drives many different research scenarios as parallel
\emph{lanes}: wet-lab biology, social-science surveys, ML systems
benchmarks, and theoretical/simulation studies. Each lane is a chain
of pluggable agent modules that the user can swap, extend, or re-wire
through ordinary YAML and through any GUI/TUI coding agent. A typed
knowledge graph below the kernel feeds every lane. The contribution of
this work is the \emph{kernel inside the machine}: a thin runtime plus
a four-field contract that lets disciplines, scenarios and discussion
modes be expressed as data rather than orchestrator code, and a
knowledge layer that lets every run accumulate into a long-lived
corpus.}
\label{fig:teaser}
\end{figure}

The past two years have produced a wave of \emph{autonomous research
systems} that aspire to take a research question --- or even a research
\emph{area} --- and return reviewed scientific output without human
intervention. Sakana AI's \textsc{AI-Scientist} v1~\cite{aisciv1}
demonstrated that GPT-4-class models~\cite{gpt4} can drive a five-stage
pipeline (idea $\rightarrow$ novelty check $\rightarrow$ experiment
$\rightarrow$ writing $\rightarrow$ review) on template tasks; v2~\cite{aisciv2}
replaced the templates with a Best-First Tree Search over hypotheses and
produced workshop-grade papers. Concurrently, PaperOrchestra~\cite{paperorchestra}
formalised a fixed five-agent writing recipe (Outline / Plotting / Lit
Review / Section Writing / Refinement) packaged as host-executable skills;
Karpathy's \emph{autoresearch} reduced neural-architecture search to a
$1{,}000$-line program-as-skill; AutoSOTA~\cite{autosota} optimised
$105$ already-published papers' code; Tongyi DeepResearch~\cite{deepresearch}
treated deep retrieval as a $30$\,B-A3B MoE end task;
InternAgent~\cite{internagent} combined generation, verification and
evolution agents for hypothesis discovery; and
ResearchAgent~\cite{researchagent} formulated iterative idea generation
over scientific literature. A parallel literature on multi-agent
collaboration~\cite{metagpt,autogen,reflexion,react} provides the building
blocks but does not target end-to-end research itself.

\paragraph{Five structural limitations.}
Read together --- with the necessary acknowledgement that no single
system has \emph{all} of these in its strongest form --- these systems
share five structural problems that motivate the present work.

\textbf{(L1) Frozen, monoculture control flow.} The shape of the research
pipeline --- how many stages, what runs in parallel, when to loop, what
form the discussion takes --- is hard-coded in Python control flow or in
a \code{WorkflowState} enum. Adding an extra review pass, swapping an LLM
call, or expressing a discipline whose loop is not ``ideate $\to$
experiment $\to$ write'' requires modifying the orchestrator. The
research \emph{process} varies enormously across disciplines and
questions: a wet-lab biology study iterates on protocols and replicates;
a social-science investigation cycles through survey design, IRB,
recruitment, coding; a CS systems paper runs benchmarks; a theoretical
physics paper proves; a meta-analysis aggregates. Discussion modes also
differ: idea-only multi-round critique before any experiment;
experiment-result-driven idea refinement; paper-result-driven post-hoc
discussion. Today's autonomous research systems target a single shape
--- typically the small-scale ML benchmark loop --- and provide no
obvious way to express the others.

\textbf{(L2) Bounded ideation, especially across domains.} LLM ideation
is fundamentally limited by the model's context window. Even the
longest-context frontier models cannot ingest, in one prompt, the
cumulative literature a human researcher carries to a cross-domain
question; and what they can ingest exhibits the well-documented
\emph{lost-in-the-middle} phenomenon~\cite{lostmiddle}. Single-call
ideators therefore see only a subsample of what would inform a real
research move, and the subsample is chosen poorly relative to the
long-tail accumulated knowledge that drives cross-domain
creativity~\cite{swansonlbd}. Researchers attempting cross-domain work
without prior accumulation in one of the domains often cannot even
formulate a starting direction.

\textbf{(L3) Episodic, non-indexed full-text reading.} Several
existing systems do load full PDFs at the moment of use:
\textsc{AI-Scientist} v1/v2~\cite{aisciv1,aisciv2} parse paper bodies
via \code{pymupdf} during novelty checking, and Tongyi
DeepResearch~\cite{deepresearch} reads full text in its information-
seeking loop. The problem is not whether the body is read \emph{once},
but that the parsed text is not retained as a queryable, typed corpus
that subsequent steps and subsequent runs can search. In the systems
above, a paper is parsed for the current call and then forgotten ---
its detailed content, figure structure, and table values are not added
to a long-lived index. PaperOrchestra~\cite{paperorchestra}, in
contrast, never reads paper bodies at all; its literature agent uses
the Semantic Scholar API which exposes only title and abstract for the
discovery step. In both regimes the system has nowhere to compensate
when full text is unavailable for a given paper, because nothing was
kept from previous reads. The right model is the researcher with deep
prior accumulation, who understands a new paper from its abstract
because the abstract resolves \emph{against} their accumulated
context.

\textbf{(L4) Code access exists, but the paper$\leftrightarrow$code link
is per-task, not corpus-level.} Several systems do touch experimental
code: AutoSOTA~\cite{autosota} explicitly optimises $105$
already-published papers' code-bases through an Idea-Library hypothesis
tracker; AI-Scientist~\cite{aisciv1,aisciv2} writes training scripts
through Aider; Karpathy's \emph{autoresearch} mutates a single
\code{train.py} under a coding-agent budget. What none of them builds
is a \emph{typed, queryable graph} that links paper concepts to
specific repository fragments and across-paper to similar repositories
from neighbouring research lines. As a result the systems cannot
answer corpus-level questions like ``find me the preprocessing pattern
this paper describes, but as it appears in a sibling paper that did
ship code'', or ``which repositories from \emph{related} research lines
implement an experimental scheme close to this newly-generated idea?''.
Across CS, ML, robotics and increasingly biology, the GitHub repository
is often the only complete specification of an experimental method ---
exact hyperparameters, preprocessing, hidden constants. The
paper$\leftrightarrow$code correspondence carries information that is
in neither the paper alone nor the code alone, and a corpus-level
index of that correspondence serves two roles: \emph{(a)} a verifier
and analyser of a paper's claims; \emph{(b)} a reservoir of inspiration
for related research whose experimental schemes are not openly
released, including for novel ideas the system itself proposes.

\textbf{(L5) No retrievable cross-run accumulation.} A human researcher's
productivity depends critically on the corpus of seeds, failed leads,
half-finished hypotheses, replication notes, and code links accumulated
across prior projects. Existing autonomous systems treat each run as an
isolated session: the crawler re-fetches papers, the ideator re-generates
seeds, the reviewer re-derives standards. Yet \emph{too much}
accumulated context overflows the model. Whether one calls the LLM once
per pipeline or many times per stage, the engineering challenge is the
same: \emph{how to surface the most useful slice of accumulated knowledge
into a finite context window at every step.} This is the central open
problem the rest of \parness{} is structured to attack.

To these five we attach two further structural observations that fall
out of (L1)--(L5) but are worth naming separately. \textbf{(L1+) Agent--
framework conflation:} domain logic
(``did this idea pass the bar?'', ``stop iterating when novelty plateaus'')
lives inside the orchestrator code, so the \emph{framework} knows about
scoring, novelty, and termination semantics it has no business knowing.
\textbf{(L1++) Coverage gaps:} most systems address a slice of the
life-cycle (only ideation; only optimisation; only retrieval; only the
writing of a draft). PaperOrchestra explicitly assumes a complete
\code{(idea, experimental log, template, guidelines)} tuple as input,
deferring everything upstream of writing to ``the host agent''.
AI-Scientist covers more stages but on hard-coded ML templates.

\paragraph{This work.}
\parness{} attacks all five (and the two corollaries) by inverting the
relationship between the \emph{framework} and the \emph{agents}. The
framework is a thin DAG runtime: $\sim\!600$ lines of \code{GraphRunner}
that does topological scheduling, input/output mapping, and process-pool
execution. Every domain decision --- whether an idea continues to the
next round, whether to fan out to parallel reviewers, when to terminate
iterative refinement --- is encoded by a module returning four reserved
fields (\S\ref{sec:contract}). On top of this kernel we ship $130{+}$
registered modules organised into $50$ YAML pipelines covering the full
life-cycle: research/crawler, full-PDF parser, code-link extractor,
ideation (six cognitive roles + twelve specialty agents), experiment
runner/verifier CLI, writing/review, and a Knowledge-Graph subsystem
($17$ adapters, $4$ pipelines, $52$ passing tests) that indexes papers,
ideas, experiments \emph{and code repositories} as typed graph nodes.
The framework deliberately exposes its modules as drop-in slots that any
GUI/TUI coding agent (Claude Code, Cursor, Copilot, OpenCode, Kilo Code)
can rewrite, replace, or compose into a custom pipeline.

\paragraph{Contributions.}
We make four contributions, mapped one-to-one with the limitations
above and elaborated as the four pillars of \S\ref{sec:method}:
\begin{enumerate}[leftmargin=1.4em,itemsep=0.2em]
\item \textbf{(M$\to$L1) A highly-customisable DAG kernel and a
declarative pipeline DSL.} A minimal four-field Agent contract
(\S\ref{sec:contract}) plus a YAML topology language with layered
validation (\S\ref{sec:dag}), so disciplines and discussion modes are
expressed as data, not code. Compared to programmatic orchestrators
(AutoGen~\cite{autogen}, MetaGPT~\cite{metagpt}, DSPy~\cite{dspy}) and
skill-pack approaches (PaperOrchestra~\cite{paperorchestra}), \parness{}
treats the \emph{topology} of the workflow as user-editable data and
enforces correctness with a layered pipeline validator
(\S\ref{sec:validation}).
\item \textbf{(M$\to$L3) A complete full-text PDF-parsing and
literature-library subsystem.} An integration of the third-party
PDF-Extract-Kit~\cite{pdfextractkit} (used unmodified except for
stability and adapter glue) feeds a typed indexing layer
(\S\ref{sec:pdfflow}); the system gracefully degrades to abstract-only
when full text is unavailable, and progressively gains
``abstract-comprehension'' capability as the cumulative corpus grows.
\item \textbf{(M$\to$L2,L4,L5) A KG-based knowledge index over
papers, ideas, experiments and code repositories} (\S\ref{sec:kg},
\S\ref{sec:codelink}). Eight-phase \textsc{Neo4j} indexer with four edge
types --- similarity, contradiction, cross-domain, derivation --- so
each LLM step retrieves the slice that fits its scenario. Six
cognitive-role idea agents (\S\ref{sec:cognitive}) consume that slice in
parallel. Five \textsc{SQLite} stores keep cross-run state durable so
the next run starts where the last left off.
\item \textbf{(M$\to$flexibility) A GUI/TUI-first extension surface.}
Every module is a single class behind a single contract, registered with
a one-line decorator, schema-validated against its YAML node, and made
discoverable to external coding agents (Claude Code, Cursor, Copilot,
OpenCode, Kilo Code) through a stable layout (\S\ref{sec:guitui}). Users
add, remove or replace modules and edit pipelines from inside their
preferred editor.
\end{enumerate}
The reference implementation is open-source: $130{+}$ modules, $50$
pipelines, $5$ \textsc{SQLite} stores, a \textsc{Neo4j} KG, with broad
development-test coverage; it runs end-to-end in single-node and
multi-GPU configurations (\S\ref{sec:eval}). Source and pipeline
configurations are released at
\url{https://github.com/gtrhythm/PARNESS}.

\section{Related Work}
\label{sec:related}

We compare \parness{} with eight recent systems that operate at adjacent
points in the design space (Table~\ref{tab:matrix}). Three axes matter for
the comparison: (i) \emph{coverage} of the research life-cycle;
(ii) \emph{composition model} for stages and decisions; and
(iii) \emph{persistence} of cross-run knowledge.

\paragraph{End-to-end pipelines.}
\textsc{AI-Scientist} v1~\cite{aisciv1} fixes a five-stage pipeline (idea,
novelty, experiment, write, review) in $\sim\!2{,}100$ LoC across five
Python files, locked to per-domain templates (e.g.\ nanoGPT, 2D-Diffusion).
v2~\cite{aisciv2} drops the template requirement by introducing a
four-phase Best-First Tree Search with VLM-in-the-loop review, but still
hard-codes the four phases. OpenAGS proposes ``folder = agent'' and a
DIRECTIVE/STATUS protocol over \code{SOUL.md} role definitions, but at
v$0.0.4$ ships placeholders for paper writing and Rust-CLI bridges.

\paragraph{Skill-pack approaches.}
\textsc{PaperOrchestra}~\cite{paperorchestra} is the most recent and
closest non-\parness{} alternative. Its design is deliberately minimal: a
five-agent recipe (Outline $\rightarrow$ Plotting $\parallel$ Lit-Review
$\rightarrow$ Section-Writing $\rightarrow$ Refinement) packaged as
\emph{skill-pack instructions} that any host coding agent (Claude Code,
Cursor, Antigravity, Cline) reads and executes using its native LLM, web
search, and shell tools. The repo ships zero embedded API clients and
zero LLM dependencies; deterministic helpers enforce citation gates,
Levenshtein matching, and halt rules. Strengths: trivially pluggable
across hosts, faithful to the published prompts, no key management.
Weaknesses for our purposes: \emph{(i)} the topology is fixed at five
steps with parallelism only between Plotting and Lit-Review; \emph{(ii)}
the entire pipeline assumes the inputs $(I,E,T,G,F)$ already exist (idea,
experimental log, template, guidelines, optional figures), so the
upstream stages of the research life-cycle (literature acquisition,
hypothesis generation, experiment design \emph{and} execution) are
explicitly out of scope or delegated to a separate optional
\code{agent-research-aggregator} skill; \emph{(iii)} cross-run knowledge
is not modelled --- the \code{provenance.json} file captures input/output
hashes for one run, not a corpus that the next run can build on; \emph{(iv)}
ideation is a single LLM call inside the host agent's window, so context
limitations (L3) hit at full strength. \parness{} and PaperOrchestra are
complementary in spirit --- one could plug PaperOrchestra's writing
sub-pipeline as a single \parness{} module --- but their architectural
commitments are different.

\paragraph{Single-stage specialists.}
Karpathy's \emph{autoresearch} is a $\sim\!1{,}000$-LOC loop that runs an
external coding agent under a $5$-minute budget to mutate a
\code{train.py}; it has no LLM, no literature, no paper output.
AutoSOTA~\cite{autosota} operates at the \emph{post}-publication boundary:
$105$ already-published papers' code-bases are optimised with an
Idea-Library hypothesis tracker; its core engine is closed-source.
DeepResearch~\cite{deepresearch} is a $30.5$\,B parameter ($3.3$\,B
active) MoE retrieval agent --- single-turn information seeking, no
experiment, no writing.

\paragraph{Generation--verification systems.}
\textsc{InternAgent}~1.5~\cite{internagent} ships eight functional
agents (Survey, Scholar, Generation, Reflection, Evolution,
MethodDevelopment, Refinement, Ranking) coordinated by an $11$-state
\code{WorkflowState} enum. ResearchAgent~\cite{researchagent} formulates
research idea generation as iterative refinement over scientific
literature. Strong on hypothesis evolution, but the \emph{paper writing}
stage is absent or thin.

\paragraph{General multi-agent and pipeline frameworks.}
AutoGen~\cite{autogen}, MetaGPT~\cite{metagpt}, ChatDev/Reflexion~\cite{reflexion}
provide multi-agent conversation as the primary abstraction; DSPy~\cite{dspy}
compiles declarative LLM \emph{programs}; LangChain composes prompt chains.
None target the research life-cycle directly. \parness{} can wrap any of
them as a single module under a four-field contract; we use AutoGen and
DSPy idioms internally for some adapters but neither is a dependency.

\begin{table*}[t]
\centering
\caption{Comparison of \parness{} with eight recent autonomous research
systems across nine axes. Numbers are module / pipeline / test counts
where reported by the system or measured from its source.
\textsc{PaperO.} = PaperOrchestra; \textsc{Intern.} = InternAgent;
\textsc{auto-r.} = Karpathy's autoresearch.}
\label{tab:matrix}
\setlength{\tabcolsep}{2.7pt}
\renewcommand{\arraystretch}{1.15}
\resizebox{\textwidth}{!}{%
\begin{tabular}{l c c c c c c c c c}
\toprule
\textbf{Axis} & \textbf{v1} & \textbf{v2} & \textbf{OpenAGS} &
\textbf{PaperO.} & \textbf{auto-r.} & \textbf{AutoSOTA} & \textbf{DeepRes.} &
\textbf{Intern.} & \textbf{\parness{}} \\
\midrule
Composition       & linear     & 4-stage    & SOUL.md   & 5-step recipe & program.md & ---     & ReAct   & 11-state    & \textbf{DAG/YAML}        \\
Pipelines         & 1          & 1          & 1         & 1            & 1          & 0       & 1       & 1           & \textbf{50}              \\
Modules           & 5 files    & $\sim$15   & 8 dirs    & 7 skills      & 3 files    & ---     & ---     & 8 cls.      & \textbf{130+}            \\
LLM providers     & multi-if   & 6+         & 4 CLI     & host LLM     & external   & 1       & 1       & 3           & \textbf{6 factory}       \\
Literature srcs.  & 1          & 1          & 2         & web+S2       & 0          & 0       & 1       & 3           & \textbf{18}              \\
Persistence       & API only   & .bib cache & FS        & provenance.json & git     & ---     & jsonl   & FS          & \textbf{5 SQLite + KG}\\
Knowledge graph   & ---        & ---        & ---       & ---          & ---        & ---     & ---     & ---         & \textbf{17 adapters}     \\
Paper writing     & template   & 2 templ.   & stub      & \textbf{strong} & ---     & ---     & ---     & ---         & 11 modules               \\
Tests             & few        & ---        & 4 files   & helpers      & 0          & ---     & ---     & ---         & \textbf{553}             \\
\bottomrule
\end{tabular}}
\end{table*}

\paragraph{Where \parness{} sits.}
\parness{} is closest in spirit to \textsc{AI-Scientist} (full life-cycle,
open-source) and \textsc{PaperOrchestra} (declarative skill-style
composition) but differs structurally on \emph{four} axes simultaneously:
(i)~the pipeline shape is data, not code or a fixed recipe; (ii)~routing
decisions live in agents, not the runner; (iii)~knowledge persists across
runs in \textsc{SQLite} and \textsc{Neo4j}, not just per-run files;
(iv)~the runtime supports multi-discipline workflow shapes (lab-loop,
survey, simulation, ablation, meta-analysis) by composition, not by
forking. To our knowledge no prior open-source system simultaneously
satisfies all four.

\section{Motivation: Why a Harness, Not a Pipeline}
\label{sec:motivation}

This section grounds the abstract limitations (L1)--(L5) of
\S\ref{sec:intro} in five concrete observations from running prior
systems and our own predecessors at scale. Each observation is mirrored
by a design move in \parness{}, summarised at the end of each
subsection and elaborated in the Method section (\S\ref{sec:method}).

\subsection{Research workflows are dynamic and discipline-specific}
\label{sec:m1}

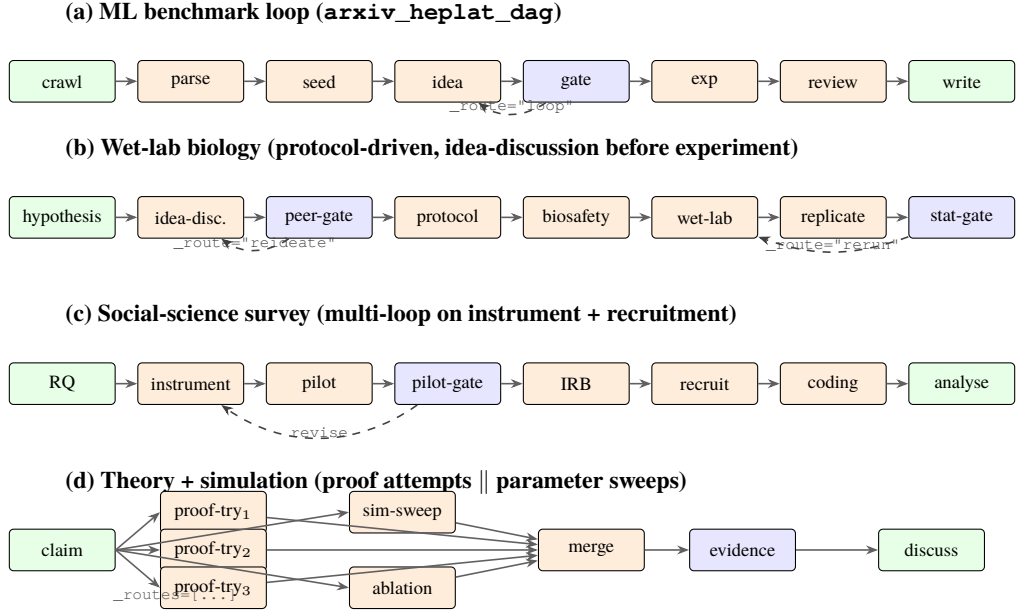
\begin{figure*}[t]
\centering
\begin{tikzpicture}[
  n/.style={draw, rounded corners=1.6pt, minimum width=14mm,
            minimum height=5.5mm, font=\scriptsize, align=center,
            inner sep=1.4pt},
  ag/.style={n, fill=orange!14},
  io/.style={n, fill=green!10},
  loop/.style={n, fill=blue!10},
  arr/.style={-{Stealth[length=1.6mm]}, semithick, gray!75!black},
  back/.style={-{Stealth[length=1.6mm]}, semithick, dashed, gray!50!black},
  hd/.style={font=\footnotesize\bfseries, anchor=west, inner sep=1pt},
]
\node[hd] (hda) at (0,2.6) {(a) ML benchmark loop (\code{arxiv\_heplat\_dag})};
\node[io] (a1) at (0,1.7)   {crawl};
\node[ag] (a2) at (1.7,1.7) {parse};
\node[ag] (a3) at (3.4,1.7) {seed};
\node[ag] (a4) at (5.1,1.7) {idea};
\node[loop](a5) at (6.8,1.7) {gate};
\node[ag] (a6) at (8.5,1.7) {exp};
\node[ag] (a7) at (10.2,1.7){review};
\node[io] (a8) at (11.9,1.7){write};
\foreach \x/\y in {a1/a2,a2/a3,a3/a4,a4/a5,a5/a6,a6/a7,a7/a8} \draw[arr] (\x) -- (\y);
\draw[back] (a5) to[bend left=35] node[font=\tiny\ttfamily,above,inner sep=0pt]{\_route="loop"} (a4);

\node[hd] (hdb) at (0,0.8) {(b) Wet-lab biology (protocol-driven, idea-discussion before experiment)};
\node[io] (b1) at (0,-0.1)   {hypothesis};
\node[ag] (b2) at (1.7,-0.1) {idea-disc.};
\node[loop](b3)at (3.4,-0.1) {peer-gate};
\node[ag] (b4) at (5.1,-0.1) {protocol};
\node[ag] (b5) at (6.8,-0.1) {biosafety};
\node[ag] (b6) at (8.5,-0.1) {wet-lab};
\node[ag] (b7) at (10.2,-0.1){replicate};
\node[loop](b8)at (11.9,-0.1){stat-gate};
\foreach \x/\y in {b1/b2,b2/b3,b3/b4,b4/b5,b5/b6,b6/b7,b7/b8} \draw[arr] (\x) -- (\y);
\draw[back] (b3) to[bend left=35] node[font=\tiny\ttfamily,above,inner sep=0pt]{\_route="reideate"} (b2);
\draw[back] (b8) to[bend left=20] node[font=\tiny\ttfamily,above,inner sep=0pt]{\_route="rerun"} (b6);

\node[hd] (hdc) at (0,-1.4) {(c) Social-science survey (multi-loop on instrument + recruitment)};
\node[io] (c1) at (0,-2.3)   {RQ};
\node[ag] (c2) at (1.7,-2.3) {instrument};
\node[ag] (c3) at (3.4,-2.3) {pilot};
\node[loop](c4)at (5.1,-2.3) {pilot-gate};
\node[ag] (c5) at (6.8,-2.3) {IRB};
\node[ag] (c6) at (8.5,-2.3) {recruit};
\node[ag] (c7) at (10.2,-2.3){coding};
\node[io] (c8) at (11.9,-2.3){analyse};
\foreach \x/\y in {c1/c2,c2/c3,c3/c4,c4/c5,c5/c6,c6/c7,c7/c8} \draw[arr] (\x) -- (\y);
\draw[back] (c4) to[bend left=35] node[font=\tiny\ttfamily,above,inner sep=0pt]{revise} (c2);

\node[hd] (hdd) at (0,-3.6) {(d) Theory + simulation (proof attempts $\parallel$ parameter sweeps)};
\node[io] (d1)  at (0,-4.5)   {claim};
\node[ag] (d2a) at (2.0,-4.0) {proof-try$_1$};
\node[ag] (d2b) at (2.0,-4.5) {proof-try$_2$};
\node[ag] (d2c) at (2.0,-5.0) {proof-try$_3$};
\node[ag] (d3a) at (4.5,-4.0) {sim-sweep};
\node[ag] (d3b) at (4.5,-5.0) {ablation};
\node[ag] (d4)  at (7.0,-4.5) {merge};
\node[loop](d5) at (9.0,-4.5) {evidence};
\node[io] (d6)  at (11.5,-4.5){discuss};
\draw[arr] (d1.east) -- ($(d2a.west)+(0,0)$);
\draw[arr] (d1.east) -- ($(d2b.west)+(0,0)$);
\draw[arr] (d1.east) -- ($(d2c.west)+(0,0)$);
\draw[arr] (d1.east) -- ($(d3a.west)+(0,0)$);
\draw[arr] (d1.east) -- ($(d3b.west)+(0,0)$);
\draw[arr] (d2a) -- (d4); \draw[arr] (d2b) -- (d4); \draw[arr] (d2c) -- (d4);
\draw[arr] (d3a) -- (d4); \draw[arr] (d3b) -- (d4);
\draw[arr] (d4)  -- (d5); \draw[arr] (d5) -- (d6);
\node[font=\tiny\ttfamily, gray!60!black] at ($(d1)+(1.5,-0.6)$) {\_routes=[\dots]};
\end{tikzpicture}
\caption{Four concrete \parness{} pipelines for four disciplines, all
expressed in the same YAML DSL on the same DAG kernel.
\textbf{(a)}~An ML benchmark loop iterates idea generation against a
quality gate, looping back to ideation if the gate score is too low.
\textbf{(b)}~A wet-lab biology pipeline holds an idea-discussion round
\emph{before} any experiment (peer-gate), and a statistical gate after
replication that can trigger reruns.
\textbf{(c)}~A survey pipeline alternates instrument-pilot-revise
loops, then sequential IRB / recruit / coding / analyse stages.
\textbf{(d)}~A theory pipeline fans out parallel proof attempts and
parameter sweeps with \code{\_routes}, then merges into an evidence
gate. Orange = LLM-driven agent; green = I/O / persistence; blue =
score-gated controller; dashed = back-edge from a routing decision.
None of these shapes can be expressed in PaperOrchestra's five-step
recipe~\cite{paperorchestra} or InternAgent's eleven-state
machine~\cite{internagent} without forking the framework.}
\label{fig:workflows}
\end{figure*}

A research \emph{process} is much more than ``ideate, experiment, write''.
A wet-lab biology study cycles through hypothesis, protocol design, IRB
or biosafety review, multiple replicates, statistical re-design, then
writing. A social-science investigation cycles through instrument
design, ethics approval, recruitment, coding, qualitative analysis,
then writing. A computer-simulation study iterates over parameter
sweeps and convergence diagnostics; a theoretical-physics paper iterates
over proofs; a meta-analysis iterates over inclusion criteria. Each of
these has \emph{its own} loop topology. Equally important, each has
\emph{its own discussion mode}: some research lines critique the idea in
multiple rounds before any experiment runs; others refine the idea
\emph{after} each experimental result lands; others run a paper-level
post-hoc discussion that drives the next paper rather than the current
one. The recent literature acknowledges this implicitly by spawning
per-discipline frameworks (Robin~\cite{robinmas} and DORA AI
Scientist~\cite{doraaisci} for biology, EXHYTE~\cite{exhyte} and
active-inference variants~\cite{activeinference} for general scientific
discovery, BioDisco for biology, GeoColab for geospatial code), but the
per-discipline framework is the wrong unit of sharing: every discipline
re-implements crawling, persistence, and review from scratch.

\paragraph{Design move (M1).} \parness{} expresses the workflow itself as
a YAML DAG over a four-field contract (Method \S\ref{sec:method:dag}).
The same kernel runs all of: a CS-systems benchmark loop, an HEP-Lat
literature-driven idea pipeline, a knowledge-graph ingestion run, and a
paper-writing-only pipeline. Adding a new discipline or a new discussion
mode is a YAML edit plus zero or more new modules registered through the
four-field contract.

\subsection{LLM ideation is bounded by context, not by talent}
\label{sec:m2}

\begin{figure}[t]
\centering
\begin{tikzpicture}[
  win/.style={draw, rounded corners=1.6pt, minimum width=20mm,
              minimum height=4.5mm, font=\scriptsize, inner sep=1pt, fill=red!8},
  win2/.style={draw, rounded corners=1.6pt, minimum width=20mm,
               minimum height=4.5mm, font=\scriptsize, inner sep=1pt, fill=blue!8},
  paper/.style={draw, rounded corners=1pt, fill=gray!10, minimum width=3mm,
                minimum height=3mm, inner sep=0pt},
  hd/.style={font=\scriptsize\bfseries, anchor=west, inner sep=1pt},
  arr/.style={-{Stealth[length=1.6mm]}, semithick, gray!75!black},
  lab/.style={font=\tiny, gray!60!black, anchor=west, inner sep=0pt},
]
\node[hd] (h1) at (0,2.4) {(a) single-call ideation: lost-in-the-middle on a fixed window};
\foreach \i in {0,...,17}{
  \node[paper, fill=gray!12] (p\i) at ($(0.4*\i,1.6)+(0,0)$) {};
}
\node[win] (w1) at (1.2,1.0) {fixed window};
\draw[gray!50, dashed] (w1.west)|-(p0.south);
\draw[gray!50, dashed] (w1.east)|-(p5.south);
\node[lab] at (3.0,1.0) {\itshape (only $\sim$5 of 18 papers fit)};
\node[lab] at (0,0.55) {\itshape attention loss in the middle of the window~\cite{lostmiddle}};

\node[hd] (h2) at (0,-0.0) {(b) \parness{}: scenario-typed retrieval into multiple cognitive roles};
\foreach \i in {0,...,17}{
  \node[paper, fill=gray!18] (q\i) at ($(0.4*\i,-0.8)+(0,0)$) {};
}
\node[font=\tiny, anchor=west] at (7.5,-0.8) {\itshape full KG corpus};

\node[win2] (r1) at (0.6,-1.7) {Reader};
\node[win2] (r2) at (2.6,-1.7) {Connector};
\node[win2] (r3) at (4.6,-1.7) {Contrarian};
\node[win2] (r4) at (6.6,-1.7) {Synthesizer};

\foreach \src/\dst in {q0/r1, q2/r1, q5/r2, q12/r2, q3/r3, q15/r3, q9/r4, q14/r4}{
  \draw[arr, gray!60] (\src.south) -- (\dst.north);
}
\node[lab] at (0,-2.4) {\itshape each role gets a typed slice tuned to its scenario (similar / cross-domain / opposite / counter-intuitive)};
\end{tikzpicture}
\caption{Cross-domain ideation under finite LLM context.
\textbf{(a)}~A single ideator with a fixed context window can only
accommodate a small subsample of the corpus, and even that subsample is
read with the well-documented attention bias of lost-in-the-middle
\cite{lostmiddle}. \textbf{(b)}~\parness{} separates retrieval from
reasoning: the KG indexer (\S\ref{sec:method:kg}) holds the full
corpus; each cognitive-role agent (\S\ref{sec:cognitive}) is wired to a
\emph{scenario-typed} retrieval (similar / cross-domain / opposite /
counter-intuitive) so the same total token budget is spent on
maximally different slices.}
\label{fig:crossdomain}
\end{figure}
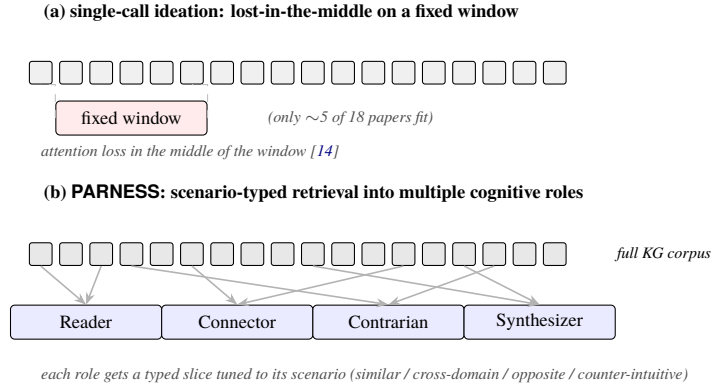

A human researcher rarely generates an idea from scratch in one sitting.
The good ideas reflect months or years of \emph{accumulated reading and
experiments}, often across disciplines. Reproducing this in an LLM is
hard for two stacked reasons. First, the long-context literature has
identified a \emph{lost-in-the-middle} pattern~\cite{lostmiddle}: even
when a frontier model technically supports a million-token window
(e.g.\ Claude Opus~4.7 with its 1\,M-token mode, DeepSeek-V4 in its
1\,M-context build, Gemini-2.x Pro at 2\,M tokens) or a quarter-million
window (e.g.\ Claude Sonnet~4.6, GPT-class long-context models), its
attention to evidence in the middle of the prompt is materially worse
than to the ends. The smaller open models that researchers actually
fine-tune and self-host (Llama-3 / Mistral / Qwen variants in the
$32$--$128$\,K range) hit the same ceiling earlier and harder. Second,
even if attention were uniform, the cumulative corpus a senior
cross-domain researcher carries (textbooks, lab notes, half-finished
papers, conference notes) does not fit in any of these windows. The
result, in practice, is that single-call ideators in autonomous
systems generate ideas that are local to whatever literature happened
to be retrieved in the last query --- a small, biased slice. The
recent \emph{IdeaSynth}~\cite{ideasynth} and
\emph{Many-Heads}~\cite{manyheads} work formalises this as a
multi-agent compositional problem.

\paragraph{Design move (M2).} \parness{} \emph{splits} ideation into
six cognitive roles
(Reader/Analyst/Connector/Contrarian/Synthesizer/Critic; \S\ref{sec:cognitive})
that each see a focused slice of the corpus retrieved through the KG and
SQLite stores. The Connector role specifically targets cross-domain
analogies; the Contrarian targets failure modes; the Synthesizer
recombines. The slice is selected by the KG retrieval adapters
(\S\ref{sec:method:kg}), which see the typed graph rather than the raw
context window. Empirically the parallel ensemble produces wider seed
coverage than a single LLM at the same total token budget; we have not
yet measured this rigorously and flag it as future work.

\subsection{Summary-only views miss the paper; cumulative reading rescues them}
\label{sec:m3:fulltext}

\begin{figure}[t]
\centering
\begin{tikzpicture}[
  pdf/.style={draw, rounded corners=1pt, fill=gray!10,
              minimum width=8mm, minimum height=11mm, font=\tiny, inner sep=1pt},
  call/.style={draw, rounded corners=1.5pt, fill=orange!12,
               minimum width=14mm, minimum height=5mm, font=\scriptsize, inner sep=1pt},
  store/.style={draw, rounded corners=2pt, fill=blue!10,
                minimum width=24mm, minimum height=8mm, font=\scriptsize, align=center, inner sep=2pt},
  arr/.style={-{Stealth[length=1.6mm]}, semithick, gray!75!black},
  trash/.style={font=\tiny, gray!50!black, anchor=west, inner sep=0pt},
  hd/.style={font=\scriptsize\bfseries, anchor=west, inner sep=1pt},
]
\node[hd] (h1) at (0,3.4) {(a) episodic full-text read (\textsc{AI-Scientist}, DeepResearch)};
\node[pdf] (pa) at (0.5,2.6) {paper};
\node[call] (ca) at (3.0,2.6) {LLM call};
\node[trash] (xa) at (5.5,2.6) {\Large $\to$ discarded};
\draw[arr] (pa) -- (ca);
\draw[arr, dashed, gray!60] (ca) -- (5.4,2.6);

\node[hd] (h2) at (0,1.6) {(b) abstract-only ideation (PaperOrchestra)};
\node[pdf, fill=gray!4] (pb) at (0.5,0.8) {abs.\,only};
\node[call] (cb) at (3.0,0.8) {LLM call};
\node[trash] (xb) at (5.5,0.8) {\Large $\to$ no fall-back};
\draw[arr] (pb) -- (cb);
\draw[arr, dashed, gray!60] (cb) -- (5.4,0.8);

\node[hd] (h3) at (0,-0.2) {(c) \parness{}: persistent index + abstract$\to$KG resolution};
\node[pdf] (pc1) at (0.4,-1.1) {paper$_1$};
\node[pdf, fill=gray!4] (pc2) at (1.5,-1.1) {abs.$_2$};
\node[pdf] (pc3) at (2.6,-1.1) {paper$_3$};
\node[store] (sc) at (5.4,-1.1) {KG index\\(typed objects)};
\node[call] (cc) at (9.5,-1.1) {LLM call};
\draw[arr] (pc1) -- (sc);
\draw[arr] (pc2) -- (sc);
\draw[arr] (pc3) -- (sc);
\draw[arr] (sc) -- (cc);
\node[font=\tiny, gray!60!black, anchor=west] at (3.5,-2.0) {abstracts are resolved \emph{against} the typed neighbourhood,};
\node[font=\tiny, gray!60!black, anchor=west] at (3.5,-2.4) {so reading capability grows monotonically with use};
\end{tikzpicture}
\caption{Reading a paper once is not the same as having it indexed.
\textbf{(a)}~\textsc{AI-Scientist}~\cite{aisciv1,aisciv2} and
DeepResearch~\cite{deepresearch} do parse full PDFs, but only
\emph{episodically}: the parsed text is consumed by the current call
and then discarded. \textbf{(b)}~PaperOrchestra~\cite{paperorchestra},
by contrast, only ever sees abstracts. In both regimes there is no
long-lived corpus that the next step or the next run can search.
\textbf{(c)}~\parness{} indexes every parsed body as typed objects
(layout, formulae, OCR, tables) into the KG; abstract-only inputs are
resolved against the typed neighbourhood, so the system's reading
capability grows monotonically with use.}
\label{fig:fulltext}
\end{figure}
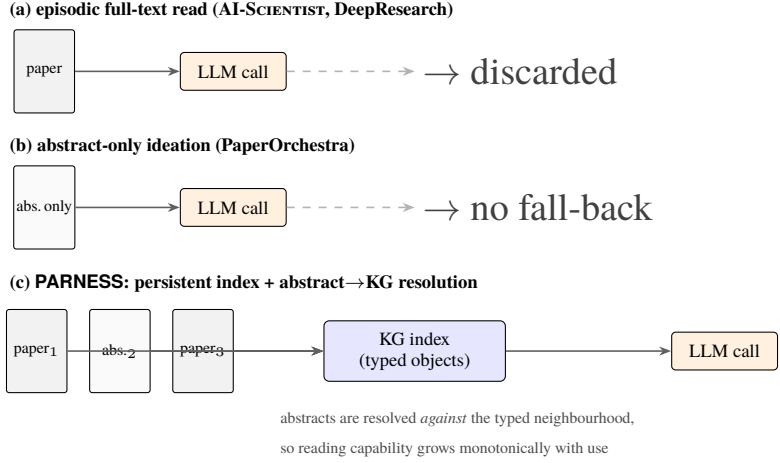

A paper's abstract is a compression of $30{-}50\times$ its body. The
contribution often hides in details that simply do not appear in the
abstract: the exact preprocessing step that made an architecture
generalise, the held-out split that exposed the bug, the table footnote
that says \emph{``three runs, mean reported''}. Several existing
systems --- AI-Scientist~\cite{aisciv1,aisciv2}, DeepResearch~\cite{deepresearch}
--- do load the full PDF when they need it, typically through a
\code{pymupdf}-style parser at the moment of use. The problem is that
the parsed text is not retained as a queryable, typed corpus. Once the
current LLM call is done, the body is forgotten, so neither a later
step in the same pipeline nor a later run can search it. Other systems
--- PaperOrchestra~\cite{paperorchestra} most prominently --- deliberately
stop at the abstract because the literature agent runs against the
Semantic Scholar API which only exposes title and abstract for the
candidate-discovery step. In both regimes the system has nowhere to
compensate when full text is unavailable, because nothing was kept from
previous reads. The correct response is therefore \emph{both} (i) ingest
the full body when it is legally and practically available, and
(ii) make every parsed body a permanent indexed asset, so subsequent
abstract-only inputs can be resolved against the typed neighbourhood.
A senior researcher in a field reads an abstract and, because of years
of accumulated context, knows roughly what the body must say. We want
\parness{} to behave the same way: a fresh installation degrades
gracefully on abstract-only input, but a \parness{} that has indexed
thousands of related papers fills in the gaps through retrieval rather
than re-reading.

\paragraph{Design move (M3).} \parness{} integrates the third-party
PDF-Extract-Kit~\cite{pdfextractkit} together with MinerU~\cite{mineru}
as the upstream parsing engines (layout, formula, OCR, tables;
\S\ref{sec:method:pdf}) and adds the surrounding plumbing to expose
the parsed objects as typed nodes to the KG indexer; the parsers
themselves are not our contribution and are used essentially
unmodified. When full text is missing, the indexer falls back to
abstract-only embedding and relies on the cross-edges built from
previously-indexed full-text neighbours. The result is a system whose
reading capability \emph{grows monotonically} with use.

\subsection{Open-source code is the missing ground truth}
\label{sec:m4:codelink}

\begin{figure}[t]
\centering
\begin{tikzpicture}[
  paper/.style={draw, rounded corners=1pt, fill=gray!12,
                minimum width=10mm, minimum height=12mm, font=\tiny,
                inner sep=1pt, align=center},
  repo/.style={draw, rounded corners=1pt, fill=orange!14,
               minimum width=11mm, minimum height=12mm, font=\tiny,
               inner sep=1pt, align=center},
  norep/.style={draw, rounded corners=1pt, fill=gray!4, dashed,
                minimum width=10mm, minimum height=12mm, font=\tiny,
                inner sep=1pt, align=center},
  derive/.style={-{Stealth[length=1.6mm]}, semithick, blue!60!black},
  inspire/.style={-{Stealth[length=1.6mm]}, semithick, dashed, gray!60!black},
  sim/.style={-, dotted, thick, orange!70!black},
  hd/.style={font=\scriptsize\bfseries, anchor=west, inner sep=1pt},
  lab/.style={font=\tiny, gray!55!black, anchor=west, inner sep=0pt},
]
\node[hd] (hp) at (-0.3,3.4) {paper nodes};
\node[paper] (P1) at (0.6,2.4) {paper\,$A$};
\node[paper] (P2) at (3.3,2.4) {paper\,$B$};
\node[norep] (P3) at (6.0,2.4) {paper\,$C$\\(no code)};
\node[paper] (P4) at (8.7,2.4) {paper\,$D$};

\node[draw, rounded corners=1pt, fill=green!14, minimum width=10mm, minimum height=6mm,
      font=\tiny, inner sep=1pt, align=center] (I) at (11.0,2.4) {new idea\\(generated)};

\node[hd] (hr) at (-0.3,0.7) {code-repository nodes};
\node[repo] (R1) at (0.6,-0.3) {repo\,$A$};
\node[repo] (R2) at (3.3,-0.3) {repo\,$B$};
\node[repo] (R4) at (8.7,-0.3) {repo\,$D$};

\draw[derive] (P1) -- node[lab,right]{derive} (R1);
\draw[derive] (P2) -- (R2);
\draw[derive] (P4) -- (R4);

\draw[sim] (R1) -- (R2);
\draw[sim] (R2) -- (R4);
\draw[sim] (R1) to[bend right=12] (R4);

\draw[inspire] (P3) to[bend right=20] node[lab,above,sloped]{cross-paper inspiration} (R2);

\draw[inspire] (I) to[bend left=15] (R4);

\node[font=\tiny, anchor=west] at (-0.3,-1.4) {\color{blue!60!black}$\rightarrow$};
\node[font=\tiny, anchor=west] at (0.0,-1.4) {derivation (paper$\to$code)};
\node[font=\tiny, anchor=west] at (3.3,-1.4) {\color{orange!70!black}\dotfill};
\node[font=\tiny, anchor=west] at (3.9,-1.4) {repository similarity};
\node[font=\tiny, anchor=west] at (7.4,-1.4) {\color{gray!60!black}-\,$-\!\rightarrow$};
\node[font=\tiny, anchor=west] at (8.4,-1.4) {inspiration / nearest-repo lookup};
\end{tikzpicture}
\caption{The \parness{} paper$\leftrightarrow$code graph. Every parsed
paper that ships a repository emits a typed \emph{derivation} edge to
its code-node; repository nodes are linked to each other by similarity
edges. A paper without code (paper~$C$) and a freshly-generated idea
both reach the closest sibling repository through cross-paper
inspiration edges. Several existing systems touch code per-task
(AutoSOTA~\cite{autosota}, AI-Scientist~\cite{aisciv1,aisciv2},
autoresearch), but none builds a corpus-level typed graph of this
shape.}
\label{fig:codelink}
\end{figure}
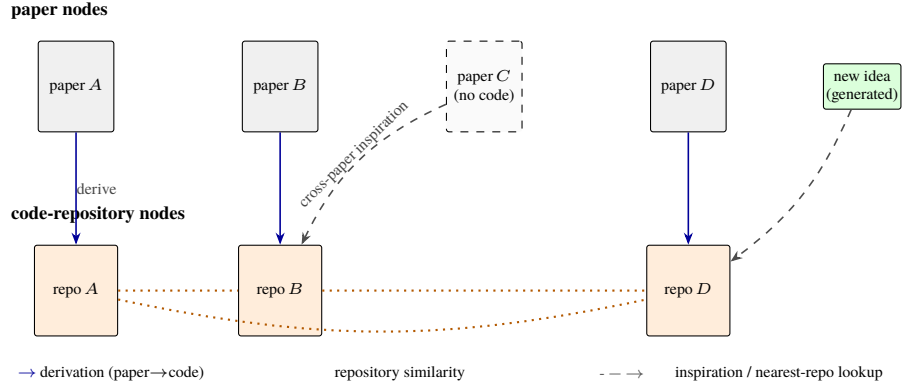

In CS and ML the open-source repository is increasingly the
\emph{contract} between author and reader. Several autonomous-research
systems already work with code: AutoSOTA~\cite{autosota} optimises
$105$ already-published papers' code-bases; Paper2Code~\cite{paper2code}
generates code from papers; AI-Scientist drives an external coding
agent (Aider) to write training scripts. The broader reproducibility
literature~\cite{leakageml,seereprod} repeatedly finds that paper-level
descriptions disagree with the code that produced the headline numbers.
What is missing is not code \emph{access}, but a typed corpus-level
\emph{graph}: in every system above, the paper$\leftrightarrow$code link
is implicit and per-task. A cross-paper retrieval like ``find me how
this paper's preprocessing is implemented in a sibling paper that
\emph{did} ship code'' is not expressible. The paper$\leftrightarrow$code
link therefore does double duty: \emph{(a)} it lets the system audit
the paper against the code, surfacing disagreements that the abstract
hides; and \emph{(b)} when a related paper does \emph{not} ship code
(common for adjacent subfields), the link to a sibling paper that
\emph{does} ship code becomes a strong source of experimental
inspiration. New ideas the system itself produces can also be indexed
against this corpus to find the closest reference implementation.

\paragraph{Design move (M4).} \parness{} ships a code-link extractor
that parses paper bodies and abstracts for repository URLs, clones
repositories under a sandbox budget, runs a code-analyser agent
to type the repository contents (training entry-point, preprocessing,
config, model definition), and writes typed nodes into the KG with
\emph{derivation} edges to their source paper. Cross-paper retrieval
adapters (\S\ref{sec:method:kg}) then surface code from related papers
when the LLM step is reasoning about a paper without code, or about a
newly-generated idea.

\subsection{Knowledge must accumulate, but accumulation must be retrievable}
\label{sec:m5}

\begin{figure}[t]
\centering
\begin{tikzpicture}[
  run/.style={draw, rounded corners=2pt, minimum width=68mm, minimum height=7mm,
              font=\scriptsize, align=left, inner sep=2pt, fill=blue!8},
  store/.style={draw, rounded corners=2pt, minimum width=18mm, minimum height=8mm,
                font=\scriptsize, align=center, inner sep=2pt, fill=orange!12},
  arr/.style={-{Stealth[length=1.6mm]}, semithick, gray!75!black},
  retr/.style={-{Stealth[length=1.6mm]}, semithick, blue!60!black, dashed},
  hd/.style={font=\scriptsize\bfseries, anchor=west, inner sep=1pt},
]
\node[hd] at (-0.3,3.0) {sequential pipeline runs over time};
\node[run] (R1) at (0,2.2) {run 1: crawl arXiv slice $\to$ ideate $\to$ experiment};
\node[run] (R2) at (0,1.4) {run 2: same idea pipeline, different RQ};
\node[run] (R3) at (0,0.6) {run 3: paper-writing-only on prior outputs};
\node[run] (R4) at (0,-0.2){run $N$: new discipline, new YAML};

\node[store] (S1) at (8.5,2.2) {SQLite\\stores};
\node[store] (S2) at (8.5,0.6) {Neo4j KG};

\foreach \r in {R1,R2,R3,R4}{ \draw[arr] (\r) -- (S1); \draw[arr] (\r) -- (S2); }

\draw[retr] (S1) to[bend left=12] (R4);
\draw[retr] (S2) to[bend left=12] (R4);

\node[hd] at (-0.3,-1.2) {finite LLM context};
\node[draw, rounded corners=1.5pt, fill=red!8, minimum width=22mm, minimum height=6mm,
      font=\scriptsize, inner sep=1pt] (W) at (3.3,-1.9) {prompt window};
\node[font=\tiny, anchor=west] at (5.0,-1.9) {{\color{gray!60!black}\itshape (single-call or multi-call: same constraint)}};
\draw[retr] (S1) to[bend right=20] node[font=\tiny, above, sloped]{retrieval per step} (W);
\draw[retr] (S2) to[bend right=18] (W);

\node[font=\tiny, anchor=west, gray!60!black] at (-0.3,-2.7)
  {\itshape every run \emph{appends} to long-lived stores; only the
   per-step retrieved slice ever enters the LLM};
\end{tikzpicture}
\caption{Knowledge accumulation across runs. Each \parness{} pipeline
appends to long-lived stores (papers, ideas, hypotheses, evidence, KG
triples). The next run begins from this corpus, retrieving only the
relevant slice for the current step. The challenge is independent of
whether the LLM is invoked once or many times: in both cases only a
finite-sized slice of accumulated knowledge can fit in the prompt, and
the engineering question is which slice.}
\label{fig:accumulation}
\end{figure}
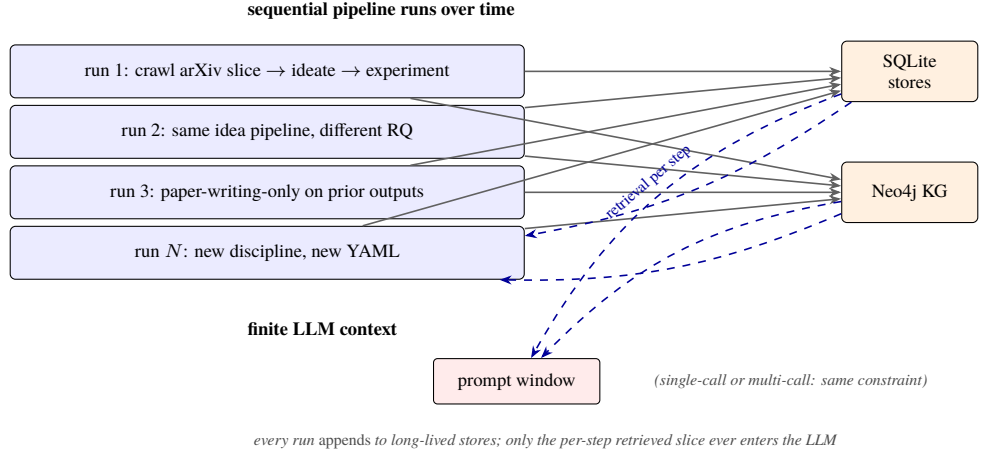

Researcher productivity depends on what was learned in earlier runs:
which literature did not pan out, which hypotheses turned out
inconsistent with experimental data, which papers from neighbouring
fields turned out useful, which code patterns reliably reproduced
results. Existing autonomous systems start each run from zero; the
result is that running the system $N$ times is no more informative than
running it once $N$ times in parallel. \emph{But} simply keeping all
prior context is also wrong, because of (L2)--(L4): you cannot re-stuff
a year of accumulated runs into one prompt, even with a million-token
window. The challenge is not single-call vs.\ multi-call --- it is
identical for both: \emph{how do we surface the smallest, highest-value
slice of the cumulative corpus into a finite context window at every
step?}

The right structure is a typed, queryable corpus that lives
\emph{outside} the LLM and is queried per-step by adapters that know
what slice this particular step needs. Recent work on hierarchical
agent memory~\cite{hierarchicalmem,gmemory,shimi} converges on the same
conclusion. Knowledge graphs offer a mature substrate for this in the
biomedical literature-based-discovery
tradition~\cite{swansonlbd,lbdsemantic}, and the recent
GraphRAG~\cite{graphrag} and TigerVector~\cite{tigervector} systems
show how to combine vector and graph retrieval over LLM-extracted
triples.

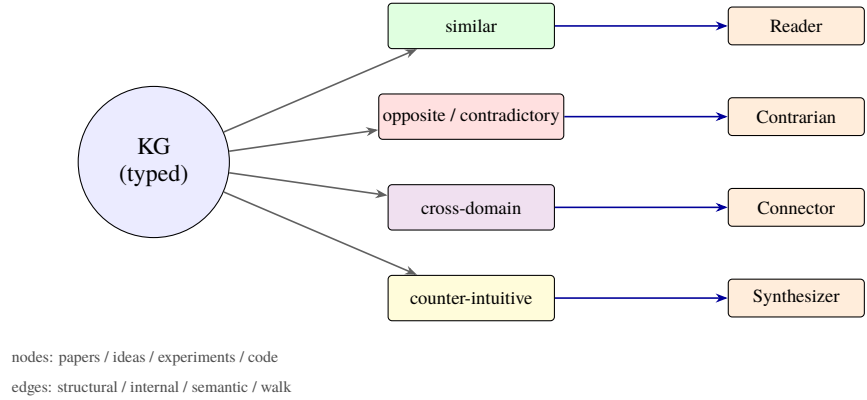
\begin{figure}[t]
\centering
\begin{tikzpicture}[
  kg/.style={draw, circle, fill=blue!8, minimum width=20mm, font=\small,
             inner sep=1pt, align=center},
  scen/.style={draw, rounded corners=1.5pt, minimum width=22mm, minimum height=6mm,
               font=\scriptsize, inner sep=1.5pt, align=center},
  role/.style={draw, rounded corners=1.5pt, fill=orange!14, minimum width=18mm,
               minimum height=5mm, font=\scriptsize, inner sep=1pt},
  arr/.style={-{Stealth[length=1.6mm]}, semithick},
]
\node[kg] (KG) at (0,0) {KG\\(typed)};

\node[scen, fill=green!12]  (s1) at (4.2, 1.8) {similar};
\node[scen, fill=red!12]    (s2) at (4.2, 0.6) {opposite / contradictory};
\node[scen, fill=violet!12] (s3) at (4.2,-0.6) {cross-domain};
\node[scen, fill=yellow!18] (s4) at (4.2,-1.8) {counter-intuitive};

\node[role] (r1) at (8.5, 1.8) {Reader};
\node[role] (r2) at (8.5, 0.6) {Contrarian};
\node[role] (r3) at (8.5,-0.6) {Connector};
\node[role] (r4) at (8.5,-1.8) {Synthesizer};

\foreach \kg/\s in {KG/s1, KG/s2, KG/s3, KG/s4}{ \draw[arr, gray!75!black] (\kg) -- (\s); }
\foreach \s/\r in {s1/r1, s2/r2, s3/r3, s4/r4}{ \draw[arr, blue!60!black] (\s) -- (\r); }

\node[font=\tiny, gray!60!black, anchor=west] at (-2.0,-2.6)
  {nodes: papers / ideas / experiments / code};
\node[font=\tiny, gray!60!black, anchor=west] at (-2.0,-3.0)
  {edges: structural / internal / semantic / walk};
\end{tikzpicture}
\caption{Scenario-typed retrieval over the \parness{} knowledge graph.
The KG holds papers, ideas, experiments and code repositories as typed
nodes connected by four edge types (structural / internal / semantic /
walk). Retrieval adapters compose those edges into four scenario
presets --- \emph{similar} (default RAG), \emph{opposite} (contradictory
results, useful for the Contrarian role), \emph{cross-domain}
(long-range walks, useful for the Connector role) and
\emph{counter-intuitive} (semantic outliers, useful for the Synthesizer
role). Each cognitive-role agent (\S\ref{sec:cognitive}) is wired to
the scenario that fits its cognitive demand, so the same finite context
window does maximally different work in each parallel slot.}
\label{fig:kgscenarios}
\end{figure}

\paragraph{Design move (M5).} \parness{} ships five \textsc{SQLite}
stores capturing orthogonal aspects of state, plus an eight-phase
\textsc{Neo4j} knowledge-graph indexer (\S\ref{sec:kg}). The graph
holds deterministic structural edges (foreign keys), LLM-discovered
internal edges (intra-batch relations), semantic edges (vector + LLM
bucket), and weighted random-walk long-range edges. The retrieval
adapters (\code{kg\_vector\_search}, \code{kg\_graph\_traverse},
\code{kg\_nl\_query}, \code{kg\_abstract\_enrich}, \code{kg\_synthesize})
are queried per-step rather than dumping the whole graph into the
prompt. Crucially, retrieval is parameterised by \emph{scenario}
(``find similar work'', ``find contradictions'', ``find cross-domain
analogies'', ``find counter-intuitive observations'') so each LLM call
gets the slice that matches its cognitive role
(Figure~\ref{fig:kgscenarios}).

\subsection{Two further pressures}
\label{sec:m6}

Beyond the five primary points we briefly note two additional pressures
that informed the design but that we do not yet evaluate empirically.

\textbf{(M$^\prime$) Reproducibility and verification.} Recent surveys
document a deep reproducibility crisis in
ML~\cite{leakageml,paper2code} and software-engineering~\cite{seereprod}
contexts. Most autonomous-research systems do not separate
\emph{running} an experiment from \emph{verifying} it. We include an
explicit verifier-augmented experiment-runner CLI (\S\ref{sec:expcli})
that produces a verifier output alongside the raw result; the verifier
output is itself a typed artefact persisted to the KG.

\textbf{(M$^{\prime\prime}$) Compositionality with existing tools.} A
research framework that cannot reuse Hugging Face Transformers for
inference, AutoGen~\cite{autogen} agents for chat, or PaperOrchestra's
writing recipe has chosen to be a walled garden. The four-field contract
(\S\ref{sec:contract}) is deliberately small precisely so that any of
these can be wrapped as a single module.

\section{Design Principles}
\label{sec:design}

Before describing the architecture we state the design principles that
drive every concrete choice in \S\ref{sec:arch}--\S\ref{sec:impl}.

\paragraph{P1 --- Framework knows nothing about the domain.}
The runner sees only nodes, edges, and a flat key/value context. It never
inspects a node's payload to decide \emph{what} the system should do
next; that decision is encoded by the upstream node returning
\code{\_route}, \code{\_routes}, or \code{\_score}.

\paragraph{P2 --- Composition is data.}
A pipeline is a YAML file. Adding a reviewer, swapping an LLM, inserting
an extra novelty filter does not require touching Python: it is a YAML
edit plus a one-line module registration.

\paragraph{P3 --- Failure is local.}
Every node runs in its own subprocess slot. GPU memory and Python state
are released by \code{os.\_exit(0)} after each node, so a leaking PyTorch
model in stage $k$ cannot poison stage $k{+}1$. Retry/timeout policy is
per-node, declared in YAML.

\paragraph{P4 --- Knowledge survives runs.}
Idea seeds, hypotheses, evidence, replication problems, transfer ideas
and KG triples persist in five normalised \textsc{SQLite} stores plus
\textsc{Neo4j}. A new pipeline run starts from the accumulated knowledge
of all prior runs --- not from scratch.

\paragraph{P5 --- LLMs are interchangeable.}
A six-provider factory (OpenAI~\cite{gpt4}, Anthropic, GLM, MiniMax, local,
mock) sits behind a single \code{LLMProvider} interface. The mock provider
ships $44$ canned-response rules so every pipeline runs offline in CI.

\paragraph{P6 --- Verification is a typed output.}
Experiment runners emit a result \emph{plus} a verifier output. Both are
typed artefacts indexed into the KG. The pipeline can route on the
verifier output (e.g.\ ``rerun if verifier flagged a non-deterministic
seed'').

\section{Method}
\label{sec:method}

This section maps the five motivating limitations of \S\ref{sec:motivation}
to four concrete method pillars. Each pillar names what \parness{}
\emph{adds} on top of the design principles of \S\ref{sec:design}; the
detailed architectural realisation of each pillar is then in
\S\ref{sec:arch}.

\begin{table}[t]
\centering
\caption{Mapping from motivation to method pillars to architectural
realisation. Each row is a self-contained design move.}
\label{tab:method-map}
\small
\setlength{\tabcolsep}{4pt}
\begin{tabular}{l l l l}
\toprule
\textbf{Motivation} & \textbf{Pillar} & \textbf{Where realised} & \textbf{Verifies} \\
\midrule
L1 (dynamic, multi-discipline) & M1 DAG kernel + DSL & \S\ref{sec:method:dag},\,\S\ref{sec:contract},\,\S\ref{sec:dag} & \S\ref{sec:eval}--unit \\
L3 (summary insufficient)      & M2 Full-text + library & \S\ref{sec:method:pdf} & qualitative \\
L2,\,L4,\,L5 (context, code, accumulation) & M3 KG + cognitive roles + code-link & \S\ref{sec:method:kg},\,\S\ref{sec:cognitive},\,\S\ref{sec:kg} & \S\ref{sec:eval}-KG \\
flexibility                    & M4 GUI/TUI surface & \S\ref{sec:guitui} & qualitative \\
\bottomrule
\end{tabular}
\end{table}

\subsection{M1 --- Highly customisable DAG kernel with a declarative pipeline DSL}
\label{sec:method:dag}

The first pillar is a general-purpose DAG scheduler tailored to the
research life-cycle. The kernel itself is small ($\sim\!600$ lines of
\code{GraphRunner}), but it is paired with a deliberately rich library
of \emph{preset agents} so that pipelines can be assembled without
writing Python. Two kinds of modules ship in the library:

\begin{itemize}[leftmargin=1.4em,itemsep=0.15em]
\item \textbf{Prompt-type agents.} An LLM call wrapped in a typed
adapter (idea-extractor, idea-generator, novelty-scorer, paper-writer,
section-reviewer, etc.). The agent contains a prompt template; the
adapter handles input/output mapping, parsing, retries, and the
\code{\_route}/\code{\_score} routing fields.
\item \textbf{Strong-functional (non-prompt) modules.} Deterministic or
heavily-hand-coded units: PDF parsing, code-repository cloning, KG
ingestion phases, vector retrieval, BibTeX writing, LaTeX compilation,
test execution. These have no LLM call, expose structured outputs, and
typically run faster and more reliably than the prompt-type agents.
\end{itemize}

The library decouples \emph{what} a stage does from \emph{when} it
runs. The same idea-generator agent appears in pipelines as varied as
\code{simple\_idea\_test}, \code{arxiv\_heplat\_dag},
\code{idea\_driven\_crawl}, and \code{iclr\_multi\_agent\_pipeline}, with
different upstream/downstream wiring in each.

\paragraph{Pipeline DSL and validation (\S\ref{sec:validation}).}
A pipeline is a YAML document with a flat list of nodes. The DSL is
layered: the schema layer rejects malformed YAML; the contract layer
checks that each node's \code{input\_mapping} keys match the upstream
\code{output\_mapping} keys; the type layer cross-checks the
\code{input\_spec}/\code{output\_spec} declared on each module against
the wiring; finally the topology layer runs a Kahn-style sort to detect
cycles and unreachable nodes. We treat the validator as a first-class
artefact: $50$ shipped pipelines and $20+$ adapter wirings are checked
in CI on every commit.

\subsection{M2 --- Full-text PDF parsing and the literature-library subsystem}
\label{sec:method:pdf}
\label{sec:pdfflow}

Most public autonomous-research systems index only abstracts. \parness{}
integrates the third-party PDF-Extract-Kit~\cite{pdfextractkit} (with
MinerU~\cite{mineru} as a fall-back) as the upstream parsing engine,
used essentially unmodified except for stability fixes and adapter
glue. The integrated pipeline extracts four typed object streams from
each PDF:

\begin{itemize}[leftmargin=1.4em,itemsep=0.15em]
\item \textbf{Layout objects} --- pages, columns, headers, sections, captions.
\item \textbf{Formula objects} --- LaTeX-recovered math, with location.
\item \textbf{OCR objects} --- text spans for non-vector documents.
\item \textbf{Table objects} --- row/column structure with cell types.
\end{itemize}

Each object carries provenance back to a \code{(paper\_id, page, bbox)}
triple. The provenance routing, the typed-object adapter layer and the
downstream KG indexer (\S\ref{sec:method:kg}) are part of \parness{};
the parsing itself is the contribution of the upstream tools. The
indexer treats these typed objects as evidence anchors when extracting
insights, so a claim ``the test set was 1\,000 held-out HEP-Lat
samples'' is grounded in the specific table it came from rather than
a free-form summary.

\paragraph{Graceful degradation when full text is unavailable.} The
parser produces a \code{coverage} flag. When full text is missing
(paywalled venue, CAPTCHA, network failure, or simply not yet
downloaded), the indexer falls back to abstract-only embedding. The
fall-back is not a failure mode --- it is the common case. The system's
``abstract-comprehension'' capability is then provided by the
cumulative corpus: if many full-text neighbours of the abstract are
already indexed, the cross-edges built from those neighbours give the
LLM step enough context to reason about the abstract without re-reading
the body. This realises the M3 design move
(Figure~\ref{fig:fulltext}).

\subsection{M3 --- KG index over papers, ideas, experiments \emph{and} code}
\label{sec:method:kg}

The third pillar is the knowledge-graph indexer. Beyond the usual
paper$\to$idea node typing, \parness{} adds two artefact types most
research-agent systems leave out: \emph{experiment runs} (raw result +
verifier output, persisted via \S\ref{sec:expcli}) and \emph{code
repositories} (typed file-tree fragments, persisted via the code-link
extractor of \S\ref{sec:codelink}). All four types share the same
edge taxonomy:

\begin{itemize}[leftmargin=1.4em,itemsep=0.15em]
\item \textbf{Structural} edges (paper $\to$ idea, paper $\to$ code,
idea $\to$ experiment) replicate \textsc{SQLite} foreign keys.
\item \textbf{Internal} edges are LLM-discovered relations within a
single ingestion batch (extracted alongside the insight nodes
themselves).
\item \textbf{Semantic} edges are vector-search candidates filtered by
an LLM bucketer in the GraphRAG~\cite{graphrag,tigervector} tradition.
\item \textbf{Walk} edges are weighted random-walk discoveries that
surface long-range cross-domain connections.
\end{itemize}

The four scenario presets --- \emph{similar} / \emph{opposite} /
\emph{cross-domain} / \emph{counter-intuitive} --- compose these edge
types into retrieval queries (Figure~\ref{fig:kgscenarios}) so each
cognitive-role agent (\S\ref{sec:cognitive}) sees the slice tuned to
its prompt. This jointly addresses (L2) by parallelising specialised
ideation, (L4) by making code repositories first-class graph citizens,
and (L5) by giving every step a retrievable view of the cumulative
corpus.

\paragraph{Code-link extractor.}
\label{sec:codelink}
A dedicated module parses every indexed paper for repository URLs
(regex against \code{github.com}, \code{gitlab.com},
\code{huggingface.co}, plus README/abstract scans).
Repositories below a configured size threshold are cloned under a
sandbox. A \code{code\_analyzer} agent types the cloned tree
(\code{train/}, \code{configs/}, \code{model/}, \code{data/}) and writes
nodes for each typed file group. \emph{Derivation} edges link the typed
file groups back to their source paper; \emph{semantic} edges then
connect those file groups to similar files in other repositories,
enabling the second use case (\S\ref{sec:m4:codelink}b): when a related
paper does not ship code, the system surfaces the closest sibling
repository as an experimental scaffold for the user or downstream
agent.

\subsection{M4 --- GUI/TUI extension surface}
\label{sec:guitui}

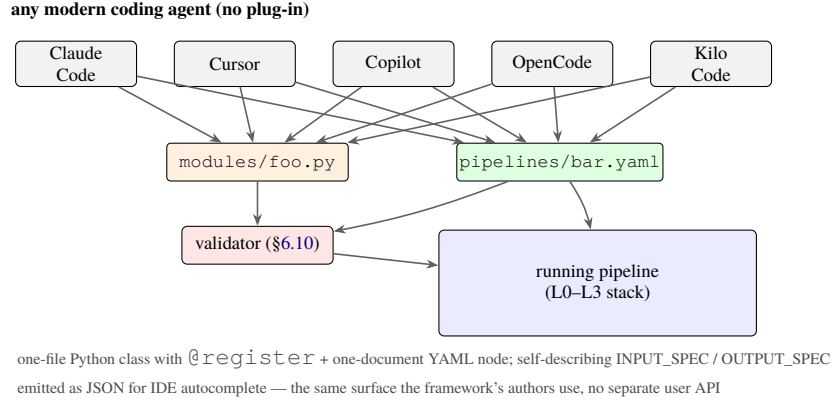
\begin{figure}[t]
\centering
\begin{tikzpicture}[
  ide/.style={draw, rounded corners=2pt, minimum width=16mm, minimum height=6mm,
              font=\scriptsize, inner sep=1.5pt, align=center, fill=gray!10},
  file/.style={draw, rounded corners=1.5pt, minimum width=24mm, minimum height=5mm,
               font=\scriptsize\ttfamily, inner sep=1pt, fill=orange!12},
  yaml/.style={draw, rounded corners=1.5pt, minimum width=24mm, minimum height=5mm,
               font=\scriptsize\ttfamily, inner sep=1pt, fill=green!12},
  dag/.style={draw, rounded corners=2pt, minimum width=42mm, minimum height=14mm,
              font=\scriptsize, inner sep=2pt, fill=blue!8, align=center},
  arr/.style={-{Stealth[length=1.6mm]}, semithick, gray!75!black},
  hd/.style={font=\scriptsize\bfseries, anchor=west, inner sep=1pt},
]
\node[hd] at (-0.3,2.6) {any modern coding agent (no plug-in)};
\node[ide] (i1) at (0.6,1.9)  {Claude\\Code};
\node[ide] (i2) at (2.7,1.9)  {Cursor};
\node[ide] (i3) at (4.8,1.9)  {Copilot};
\node[ide] (i4) at (6.9,1.9)  {OpenCode};
\node[ide] (i5) at (9.0,1.9)  {Kilo\\Code};

\node[file] (py)   at (3.0,0.6) {modules/foo.py};
\node[yaml] (yml)  at (7.0,0.6) {pipelines/bar.yaml};

\foreach \i in {i1,i2,i3,i4,i5}{ \draw[arr] (\i) -- (py); \draw[arr] (\i) -- (yml); }

\node[draw, rounded corners=2pt, minimum width=20mm, minimum height=5mm,
      font=\scriptsize, inner sep=1.5pt, fill=red!10] (val) at (3.0,-0.5) {validator (\S\ref{sec:validation})};
\node[dag] (D) at (7.5,-1.0) {running pipeline\\(L0--L3 stack)};

\draw[arr] (py) -- (val);
\draw[arr] (yml) to[bend left=5] (val);
\draw[arr] (val) -- (D);
\draw[arr] (yml) to[bend left=10] (D);

\node[font=\tiny, gray!60!black, anchor=west] at (-0.3,-2.0)
  {one-file Python class with \code{@register} + one-document YAML node;
   self-describing INPUT\_SPEC / OUTPUT\_SPEC};
\node[font=\tiny, gray!60!black, anchor=west] at (-0.3,-2.4)
  {emitted as JSON for IDE autocomplete --- the same surface the framework's
   authors use, no separate user API};
\end{tikzpicture}
\caption{Every \parness{} module is a single Python class behind a
single contract, registered with a one-line decorator and described by
a YAML node. External coding agents --- Claude Code, Cursor, Copilot,
OpenCode, Kilo Code --- can therefore add a new module, edit an
existing module, or re-wire a pipeline by editing one Python file plus
one YAML file. The pipeline validator checks the edit before the next
run starts. This is the same surface the framework's authors use; there
is no separate ``user'' API.}
\label{fig:flexibleide}
\end{figure}

The fourth pillar is a deliberately small extension surface so that a
modern GUI/TUI coding agent can act as the user-facing IDE for
\parness{} without a custom plug-in. Three properties make this work:

\begin{itemize}[leftmargin=1.4em,itemsep=0.15em]
\item \textbf{Single-file modules.} A module is one Python file with
one class subclassing \code{BaseModule} and a one-line
\code{@register("module\_name")} decorator. Adding or replacing a
module is a single-file diff.
\item \textbf{Single-document pipelines.} A pipeline is one YAML file.
Re-shaping the workflow does not touch Python at all.
\item \textbf{Self-describing schemas.} Every module declares its
\code{INPUT\_SPEC} and \code{OUTPUT\_SPEC} as Python type stubs that the
validator (\S\ref{sec:validation}) can introspect; the schemas are
emitted as JSON for IDE autocomplete.
\end{itemize}

The result is that the same workflow we use during development ---
``ask Claude Code to add a new \code{kg\_random\_walk\_v2} module that
samples by node degree'' --- works for any user with any of the listed
coding agents. We do not ship a custom plug-in; the IDE is whatever
coding agent the user already trusts.

\section{System Architecture}
\label{sec:arch}

\subsection{Layered overview}
\begin{figure}[t]
\centering
\begin{tikzpicture}[
  band/.style={draw, rounded corners=3pt, minimum height=18mm,
               minimum width=92mm, font=\small, align=left, inner sep=4pt,
               text width=82mm},
  l0style/.style={band, fill=blue!8},
  l1style/.style={band, fill=gray!12},
  l2style/.style={band, fill=orange!12},
  l3style/.style={band, fill=green!12},
  arrow/.style={-{Stealth[length=2.5mm]}, thick, gray!70},
  tag/.style={font=\footnotesize\bfseries, anchor=east, inner sep=1pt},
  side/.style={font=\scriptsize\itshape, gray!60!black, anchor=west, inner sep=1pt, align=left},
  node distance=2mm,
]
\node[l3style]                                   (l3)
  {\textbf{Pipeline} \quad {\small\itshape user surface}\\
   \;$\bullet$\; YAML pipeline DSL ($\sim$50 shipped pipelines)\\
   \;$\bullet$\; Entry-point scripts users run from the shell\\
   \;$\bullet$\; Pipeline validator (schema / contract / type / topology)};
\node[l2style, below=of l3]                      (l2)
  {\textbf{Agents} \quad {\small\itshape research workers}\\
   \;$\bullet$\; 25+ domain agents (idea, paper, experiment, KG, review)\\
   \;$\bullet$\; 130+ adapters wrapping agents \& tools as \code{BaseModule}\\
   \;$\bullet$\; LLM provider factory (6 + mock); domain tools (PDF, KG,\\
   \quad\;\, crawlers, experiment-runner CLI)};
\node[l1style, below=of l2]                      (l1)
  {\textbf{Persistence} \quad {\small\itshape durable state}\\
   \;$\bullet$\; 5 \textsc{SQLite} stores (papers, knowledge,\\
   \quad\;\, evaluations, paper-writing, experiments)\\
   \;$\bullet$\; \textsc{Neo4j} knowledge graph + native vector index\\
   \;$\bullet$\; Monitoring, exception, run-context utilities};
\node[l0style, below=of l1]                      (l0)
  {\textbf{DAG kernel} \quad {\small\itshape scheduler + contract}\\
   \;$\bullet$\; \code{GraphRunner} (topological scheduling, $\sim$600 LoC)\\
   \;$\bullet$\; \code{BaseModule} four-field Agent contract\\
   \;$\bullet$\; \code{ModuleRegistry}; process-pool isolation};

\foreach \a/\b in {l3/l2, l2/l1, l1/l0}{ \draw[arrow] (\a) -- (\b); }

\node[tag] at ($(l3.west)+(-2mm,0)$) {L3};
\node[tag] at ($(l2.west)+(-2mm,0)$) {L2};
\node[tag] at ($(l1.west)+(-2mm,0)$) {L1};
\node[tag] at ($(l0.west)+(-2mm,0)$) {L0};

\node[side] at ($(l3.east)+(2mm,0)$) {what the user writes};
\node[side] at ($(l2.east)+(2mm,0)$) {what does the work};
\node[side] at ($(l1.east)+(2mm,0)$) {what survives a run};
\node[side] at ($(l0.east)+(2mm,0)$) {what schedules everything};

\end{tikzpicture}
\caption{The four-layer architecture of \parness{}. \textbf{L0 (DAG
kernel)} is the foundation: a thin scheduler plus a four-field contract.
\textbf{L1 (Persistence)} keeps state durable across nodes and across
runs. \textbf{L2 (Agents)} is the population of LLM workers and tools
that do the actual research. \textbf{L3 (Pipeline)} is the user surface:
YAML pipelines and entry-point scripts. Configuration flows downward at
start time; data and persisted knowledge flow upward at runtime.}
\label{fig:layers}
\end{figure}
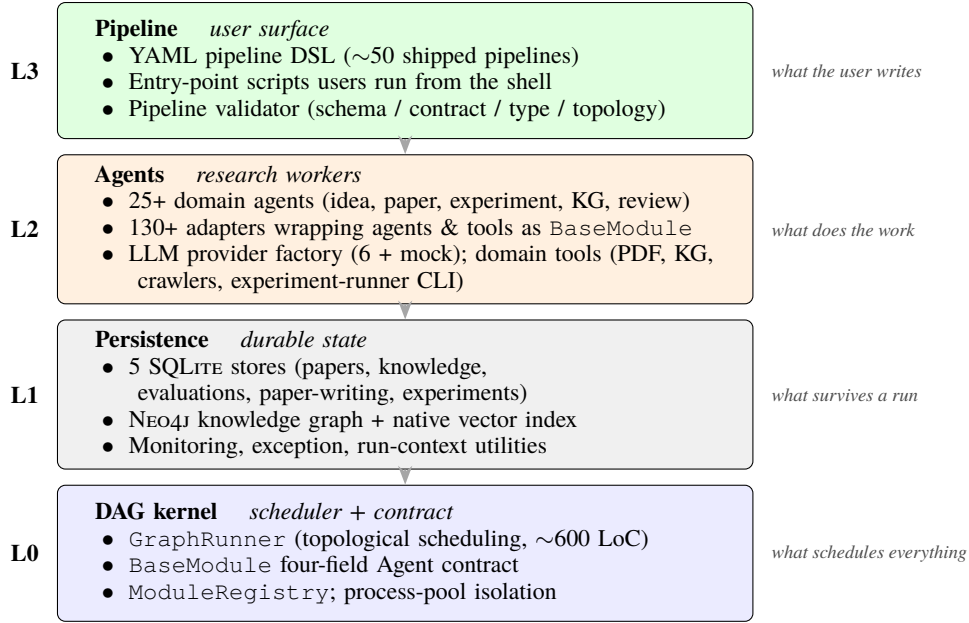

\parness{} is organised into four layers (Figure~\ref{fig:layers}).
\textbf{L0, the DAG kernel,} is the foundation everything else stands on:
\code{GraphRunner} (topological scheduling, $\sim\!600$ LoC), the
four-field \code{BaseModule} contract (\S\ref{sec:contract}), and the
\code{ModuleRegistry}. \textbf{L1, persistence,} provides durable state
that survives a single node's lifetime: five \textsc{SQLite} stores
plus a \textsc{Neo4j} knowledge graph with a native vector index, and
shared monitoring / exception / run-context utilities. \textbf{L2,
agents,} is the population of research workers: $25{+}$ LLM-driven
domain agents (idea, paper, experiment, KG, review), $130{+}$ adapters
wrapping agents and tools as \code{BaseModule}, the LLM provider
factory (six providers + mock), and the domain-tool layer (a
PDF-parsing wrapper around the external PDF-Extract-Kit~\cite{pdfextractkit}
service, KG, crawlers, experiment-runner CLI). \textbf{L3, pipeline,} is the user surface:
$\sim\!50$ shipped YAML pipelines plus the entry-point scripts users
run from the shell, gated by a layered pipeline validator
(\S\ref{sec:validation}).

\subsection{The Agent contract}
\label{sec:contract}
The contract between the runner and a module is deliberately small. A
module subclasses \code{BaseModule} and implements one method:
\code{async execute(inputs: Dict) -> Dict}. The runner inspects the
returned dictionary for four reserved fields:

\begin{lstlisting}[caption={The four-field Agent contract}, label={lst:contract}]
{
  "_route":    "next_node_id",      # single successor
  "_routes":   ["a", "b", "c"],     # one-to-many fan-out
  "_score":    0.83,                # numeric for iteration / ranking
  "_metadata": {...},               # provenance, free-form
  ...                               # plus all domain outputs
}
\end{lstlisting}

The runner removes these four keys before passing the rest to downstream
nodes. The semantics are simple: if \code{\_route} is set the runner
follows it; if \code{\_routes} is set it fans out; if \code{\_score} is
set the value is recorded for upstream iteration controllers; otherwise
the runner falls back to standard topological successors.

This simplicity has three consequences. First, \emph{any} module can be a
routing decision: a quality-scoring agent emits
\code{\_route = "continue"} or \code{"stop"} based on its own threshold
logic. Second, fan-out is ordinary: an idea-generator returns
\code{\_routes} with one entry per generated idea and the runner
instantiates one experiment node per entry. Third, iteration is not
special: a controller module reads the latest \code{\_score} and emits
\code{\_route = "loop"} or \code{"exit"}, which is indistinguishable from
any other branching decision.

\subsection{Declarative pipeline DSL}
\label{sec:dag}

A pipeline is a YAML document with three top-level keys: \code{nodes},
\code{edges}, and \code{config}. \code{nodes} is the only one that
matters for execution; \code{edges} is purely declarative (used for
documentation and visualisation). Each node carries the standard fields:
\code{id}, \code{module}, \code{depends\_on}, \code{params},
\code{input\_mapping}, \code{output\_mapping}, \code{routes},
\code{timeout}, and \code{retry}. Listing~\ref{lst:yaml} shows a six-node
fragment from \code{auto\_research.yaml}.

\begin{lstlisting}[caption={Pipeline fragment showing node + adapter wiring}, label={lst:yaml}]
nodes:
  - id: extract
    module: idea_extractor
    input_mapping:
      papers: paper_store.papers
    output_mapping:
      seeds: extract.seeds
  - id: generate
    module: idea_generator
    depends_on: [extract]
    input_mapping:
      seeds: extract.seeds
  - id: gate
    module: quality_scorer
    depends_on: [generate]
    routes:
      "continue": evaluate
      "stop":     export
config:
  max_rounds: 100
  max_parallel: 0
\end{lstlisting}

\begin{figure}[t]
\centering
\begin{tikzpicture}[
  nd/.style={draw, rounded corners=2pt, minimum width=18mm,
             minimum height=7mm, font=\small, align=center, inner sep=2pt},
  agent/.style={nd, fill=orange!12},
  next/.style={nd, fill=green!10},
  ctrl/.style={nd, fill=blue!8, font=\footnotesize},
  arr/.style={-{Stealth[length=2mm]}, thick},
  lab/.style={font=\scriptsize\ttfamily, inner sep=1pt},
  panel/.style={font=\footnotesize\bfseries\itshape, anchor=west, inner sep=1pt},
  sep/.style={dashed, gray!40, thin},
]

\node[panel] at (-0.3, 2.7) {(1) explicit successor};
\node[agent] (a1)  at (0.6, 1.7) {agent};
\node[next]  (n1a) at (5.0, 2.2) {step A};
\node[next]  (n1b) at (5.0, 1.2) {step B};
\draw[arr] (a1) -- node[lab, above, sloped, pos=0.55] {\_route="A"} (n1a);
\draw[arr, dashed, gray] (a1) -- node[lab, below, sloped, pos=0.55] {(not taken)} (n1b);

\draw[sep] (-0.6, 0.6) -- (8.5, 0.6);

\node[panel] at (-0.3, 0.3) {(2) one-to-many fan-out};
\node[agent] (a2)  at (0.6, -0.9) {generator};
\node[next]  (n2a) at (5.0,  -0.1) {worker $w_1$};
\node[next]  (n2b) at (5.0,  -0.9) {worker $w_2$};
\node[next]  (n2c) at (5.0,  -1.7) {worker $w_3$};
\draw[arr] (a2) -- (n2a);
\draw[arr] (a2) -- (n2b);
\draw[arr] (a2) -- (n2c);
\node[lab, anchor=west] at (1.9, -2.2) {\_routes=[w1,w2,w3]};

\draw[sep] (-0.6, -2.7) -- (8.5, -2.7);

\node[panel] at (-0.3, -3.0) {(3) iteration via score gate};
\node[agent] (a3)  at (0.6, -4.2) {scorer};
\node[ctrl]  (c3)  at (3.7, -4.2) {iter ctrl};
\node[next]  (b3a) at (6.7, -3.7) {refine};
\node[next]  (b3b) at (6.7, -4.7) {exit};
\draw[arr] (a3) -- node[lab, above] {\_score=0.83} (c3);
\draw[arr] (c3) -- node[lab, above, sloped, pos=0.55] {\_route="loop"} (b3a);
\draw[arr, dashed, gray] (c3) -- node[lab, below, sloped, pos=0.55] {else} (b3b);

\end{tikzpicture}
\caption{Three uses of the four-field Agent contract. The runner makes
no domain decisions: it merely follows the routing fields the upstream
module emits.}
\label{fig:contract}
\end{figure}
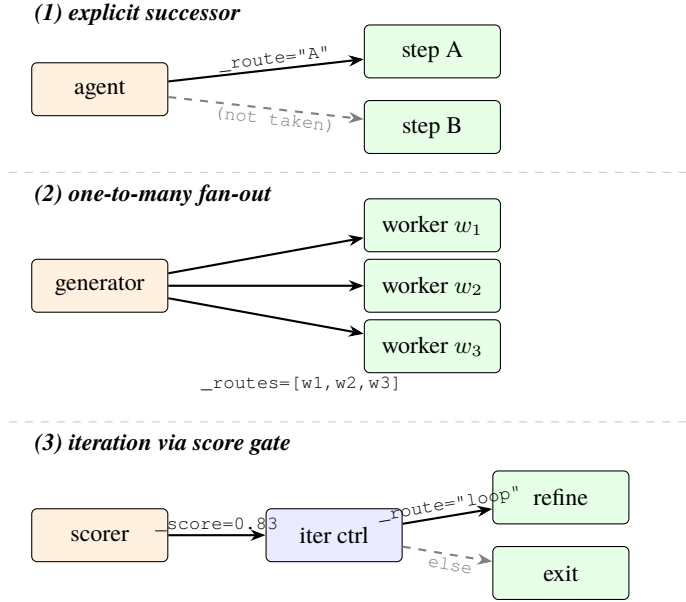

\code{GraphRunner} chooses among three scheduling strategies based on
what each node returns (Figure~\ref{fig:contract}).
(1)~If a node returns \code{\_route}, the runner takes the explicit
successor named in \code{routes[\_route]}.
(2)~If \code{\_routes} is set the runner clones the downstream sub-DAG
once per element. (3)~Otherwise it falls back to a Kahn-style
topological sort that combines \code{depends\_on} with implicit
dependencies extracted from \code{input\_mapping} (i.e.\ if node $B$
maps from \code{output.A.field}, then $B$ depends on $A$ even if not
explicitly declared).

The runner has two backends: an in-process \code{asyncio} loop for
testing, and a \code{ProcessPoolExecutor} backend for production. The
production backend instantiates worker processes lazily and re-uses them
across nodes, but enforces \code{os.\_exit(0)} after each node's work to
release GPU memory.

\subsection{From fixed recipe to data-defined topology}
\label{sec:recipe}

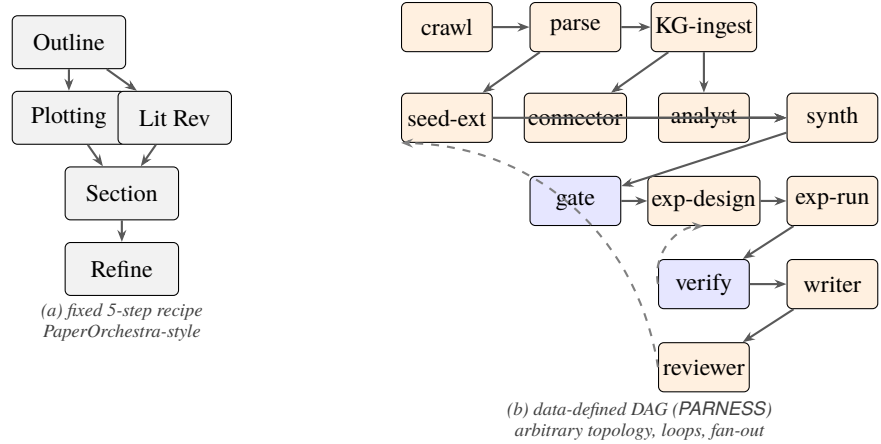
\begin{figure}[t]
\centering
\begin{tikzpicture}[
  fixed/.style={draw, rounded corners=2pt, minimum width=15mm, minimum height=7mm,
                font=\footnotesize, align=center, fill=gray!10, inner sep=1.5pt},
  dyn/.style={draw, rounded corners=2pt, minimum width=12mm, minimum height=6.5mm,
              font=\footnotesize, align=center, fill=blue!10, inner sep=1.5pt},
  agent/.style={draw, rounded corners=2pt, minimum width=12mm, minimum height=6.5mm,
                font=\footnotesize, align=center, fill=orange!12, inner sep=1.5pt},
  arr/.style={-{Stealth[length=1.6mm]}, thick, gray!70!black},
  ann/.style={font=\scriptsize\itshape, gray!50!black, align=center},
]

\node[fixed] (f1) at (0,3) {Outline};
\node[fixed] (f2) at (0,2) {Plotting};
\node[fixed] (f3) at (1.4,2) {Lit Rev};
\node[fixed] (f4) at (0.7,1) {Section};
\node[fixed] (f5) at (0.7,0) {Refine};

\draw[arr] (f1) -- (f2);
\draw[arr] (f1) -- (f3);
\draw[arr] (f2) -- (f4);
\draw[arr] (f3) -- (f4);
\draw[arr] (f4) -- (f5);

\node[ann] at (0.7,-0.7) {(a) fixed 5-step recipe\\PaperOrchestra-style};

\begin{scope}[xshift=5cm]
\node[agent] (g1) at (0,3.2) {crawl};
\node[agent] (g2) at (1.7,3.2) {parse};
\node[agent] (g3) at (3.4,3.2) {KG-ingest};
\node[agent] (g4) at (0,2.0) {seed-ext};
\node[agent] (g5) at (1.7,2.0) {connector};
\node[agent] (g6) at (3.4,2.0) {analyst};
\node[agent] (g7) at (5.1,2.0) {synth};
\node[dyn]   (g8) at (1.7,0.9) {gate};
\node[agent] (g9) at (3.4,0.9) {exp-design};
\node[agent] (g10)at (5.1,0.9) {exp-run};
\node[dyn]   (g11)at (3.4,-0.2){verify};
\node[agent] (g12)at (5.1,-0.2){writer};
\node[agent] (g13)at (3.4,-1.3){reviewer};

\draw[arr] (g1) -- (g2); \draw[arr] (g2) -- (g3);
\draw[arr] (g2) -- (g4); \draw[arr] (g3) -- (g5); \draw[arr] (g3) -- (g6);
\draw[arr] (g4) -- (g7); \draw[arr] (g5) -- (g7); \draw[arr] (g6) -- (g7);
\draw[arr] (g7) -- (g8);
\draw[arr] (g8) -- (g9); \draw[arr] (g9) -- (g10); \draw[arr] (g10) -- (g11);
\draw[arr] (g11) -- (g12); \draw[arr] (g12) -- (g13);
\draw[arr, dashed, gray] (g11.west) to[bend left=50] (g9.south);
\draw[arr, dashed, gray] (g13.west) to[bend right=40] (g4.south west);

\node[ann] at (2.55,-2.0) {(b) data-defined DAG (\parness{})\\arbitrary topology, loops, fan-out};
\end{scope}
\end{tikzpicture}
\caption{(a) PaperOrchestra-style fixed five-step recipe: writing the
paper assumes the inputs already exist, the topology is hard-coded, no
upstream stages. (b) \parness{} pipeline: composition is data, dynamic
fan-out (Connector/Analyst/etc.\ in parallel), score-gated loops (dashed
arrows back to earlier stages), and full life-cycle coverage from crawl
to review. Both diagrams represent real shipped pipelines.}
\label{fig:recipe-vs-dag}
\end{figure}

Figure~\ref{fig:recipe-vs-dag} illustrates the difference between a
fixed-recipe approach and \parness{}'s data-defined topology side by
side. The recipe approach makes individual steps powerful (each
PaperOrchestra agent has a long verbatim prompt with embedded
deterministic gates), but cannot express, e.g., a \emph{conditional}
re-execution of literature search if the section writer flags missing
citations, or a \emph{fan-out} of plotting that depends on how many
distinct experiments the experimental log contains. \parness{} handles
both as ordinary YAML (an extra route, or a \code{\_routes} return).

\subsection{Module taxonomy}
\label{sec:modules}

The $130{+}$ registered modules group into seven categories
(Table~\ref{tab:modules}). The largest category is \emph{ideation}: in
addition to the standard extractor / generator / evaluator chain it
includes six \emph{cognitive-role} agents and twelve specialty agents
(replication, transfer, critique, theory, meta-analysis, follow-up,
adversarial, limitation, hypothesis, evidence, paper-code analyser,
paper-code retrieval).

\begin{table}[t]
\centering
\caption{The $130{+}$ modules, by category. Counts grew from $116$ at the
abstract-snapshot date as the experiment-runner / verifier CLI and KG
adapters landed.}
\label{tab:modules}
\small
\begin{tabular}{lr}
\toprule
\textbf{Category} & \textbf{\# Modules} \\
\midrule
Research / crawler / parser   & 8 \\
Ideation                      & 26 \\
Experiment (incl.\ runner / verifier CLI) & 24 \\
Writing \& review             & 11 \\
Knowledge graph               & 17 \\
Infrastructure / persistence  & 16+ \\
Iteration controllers \& gates & 7 \\
Post-processing / export      & 2 \\
\midrule
\textbf{Total} & \textbf{130+} \\
\bottomrule
\end{tabular}
\end{table}

\subsection{Cognitive-role ideation}
\label{sec:cognitive}

\begin{figure}[t]
\centering
\begin{tikzpicture}[
  src/.style={draw, rounded corners=2pt, fill=blue!8, minimum width=24mm,
              minimum height=8mm, font=\scriptsize, inner sep=1.5pt, align=center},
  role/.style={draw, rounded corners=2pt, fill=orange!14, minimum width=22mm,
               minimum height=5.5mm, font=\scriptsize, inner sep=1pt, align=center},
  agg/.style={draw, rounded corners=2pt, fill=green!12, minimum width=22mm,
              minimum height=8mm, font=\scriptsize, inner sep=1.5pt, align=center},
  arr/.style={-{Stealth[length=1.6mm]}, semithick, gray!75!black},
  arrlbl/.style={font=\tiny\itshape, fill=white, inner sep=1pt, gray!60!black},
  capt/.style={font=\tiny, gray!60!black, anchor=west, align=left, inner sep=1pt},
]
\node[src] (G) at (0,0) {KG retrieval\\\code{\_routes=[\dots]}};

\node[role] (Re) at (7.5, 2.4) {Reader};
\node[role] (An) at (7.5, 1.45) {Analyst};
\node[role] (Co) at (7.5, 0.5) {Connector};
\node[role] (Ct) at (7.5,-0.5) {Contrarian};
\node[role] (Sy) at (7.5,-1.45) {Synthesizer};
\node[role] (Cr) at (7.5,-2.4) {Critic};

\draw[arr] (G.east) -- node[arrlbl, pos=0.55] {similar}        (Re.west);
\draw[arr] (G.east) -- node[arrlbl, pos=0.55] {mechanism}      (An.west);
\draw[arr] (G.east) -- node[arrlbl, pos=0.55] {cross-domain}   (Co.west);
\draw[arr] (G.east) -- node[arrlbl, pos=0.55] {opposite}       (Ct.west);
\draw[arr] (G.east) -- node[arrlbl, pos=0.55] {counter-int.}   (Sy.west);
\draw[arr] (G.east) -- node[arrlbl, pos=0.55] {filter}         (Cr.west);

\node[agg] (A) at (12.5,0) {result\\aggregator};
\foreach \r in {Re,An,Co,Ct,Sy,Cr}{ \draw[arr] (\r.east) -- (A.west); }

\node[capt] at (-0.5,-3.5)
  {fan-out via \code{\_routes}; parallel execution; deduplicated convergence};
\end{tikzpicture}
\caption{Six cognitive-role agents used in \parness{} ideation. The KG
retrieval step (left) emits a \code{\_routes} fan-out so the runner
schedules each role in parallel; each role's prompt is engineered for
one orthogonal cognitive demand and is wired to a different scenario-
typed retrieval slice (right of each role). The aggregator deduplicates
and ranks the seeds before passing them to a downstream gate. None of
the framework's domain knowledge sits inside \code{GraphRunner}: the
roles, the retrieval scenarios, and the aggregator are all ordinary
\code{BaseModule} instances.}
\label{fig:cognitive}
\end{figure}

Most prior systems use a single ``idea generator'' LLM call. Recent work
(\textsc{Many-Heads}~\cite{manyheads},
\textsc{IdeaSynth}~\cite{ideasynth}, \textsc{BioDisco}) shows that
multi-agent ideation produces more diverse and higher-quality candidates
than a single LLM at matched budget. \parness{} replaces the single call
with six agents whose prompts make orthogonal cognitive demands
(Figure~\ref{fig:cognitive}). The roles run in parallel under a fan-out
node and converge through a result-aggregator. Empirically the parallel
ensemble produces wider seed coverage than a single LLM at the same
total token budget; we have not yet measured this rigorously and flag it
as future work.

\subsection{Knowledge Graph subsystem}
\label{sec:kg}

\begin{figure*}[t]
\centering
\begin{tikzpicture}[
  phase/.style={draw, rounded corners=2pt, minimum width=20mm, minimum height=8mm,
                font=\small, align=center, inner sep=2pt},
  llm/.style={phase, fill=orange!18},
  det/.style={phase, fill=blue!10},
  arrow/.style={-{Stealth[length=2mm]}, thick, gray!70!black},
  legend/.style={font=\scriptsize, inner sep=1pt},
  node distance=4mm,
]
\node[llm] (p1)                    {1.~extract};
\node[det, right=of p1]   (p2)     {2.~dedup};
\node[det, right=of p2]   (p3)     {3.~embed};
\node[det, right=of p3]   (p4)     {4.~write\_node};
\node[llm, below=10mm of p4] (p5)  {5.~internal\_edge};
\node[det, left=of p5]    (p6)     {6.~struct\_edge};
\node[llm, left=of p6]    (p7)     {7.~semantic\_edge};
\node[llm, left=of p7]    (p8)     {8.~random\_walk};

\foreach \a/\b in {p1/p2,p2/p3,p3/p4}{ \draw[arrow] (\a) -- (\b); }
\draw[arrow] (p4) -- ++(0,-5mm) -| (p5);
\draw[arrow] (p5) -- (p6);
\draw[arrow] (p6) -- (p7);
\draw[arrow] (p7) -- (p8);

\node[legend, draw, fill=orange!18, rounded corners=1pt, anchor=north west]
  at ($(p8.south west)+(0,-7mm)$) (k1) {LLM-driven phase};
\node[legend, draw, fill=blue!10, rounded corners=1pt,
      right=2mm of k1] (k2) {deterministic phase};
\node[legend, anchor=west, right=4mm of k2, text width=85mm, align=left]
  {Output: \textsc{Neo4j} nodes + 4 edge types
   (struct / internal / semantic / walk),
   each backed by a vector index for retrieval.};
\end{tikzpicture}
\caption{The eight-phase Knowledge-Graph indexing pipeline. LLM phases
(orange) handle the open-ended steps --- extracting insights, discovering
intra-batch relations, semantic edge filtering, and weighted random-walk
relation discovery. Deterministic phases (blue) handle dedup, embedding,
persistence, and structural-edge replication from \textsc{SQLite}. The
pipeline is incremental: a new ingestion batch only re-runs phases 1--4
for new content.}
\label{fig:kg}
\end{figure*}
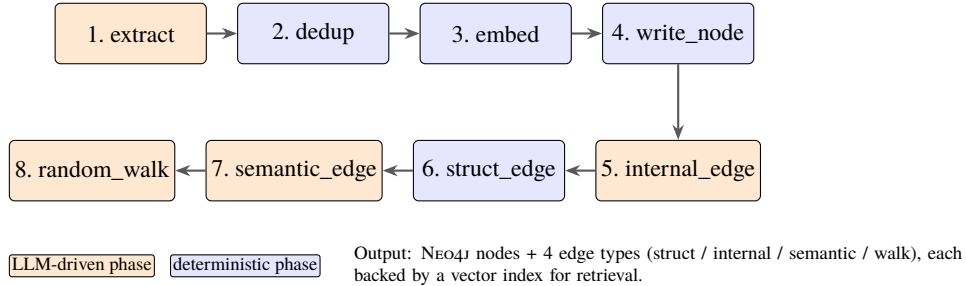

The Knowledge Graph (KG) subsystem, added in May 2026, persists every
extracted insight as a typed node in \textsc{Neo4j} and links nodes
along four edge types: \emph{structural} (deterministic foreign-key
edges from \textsc{SQLite}), \emph{internal} (LLM-discovered relations
inside an ingestion batch), \emph{semantic} (vector-search-filtered
candidate relations with LLM bucketing in the GraphRAG~\cite{graphrag}
tradition), and \emph{walk} (long-range relations discovered by weighted
random walk). Eight indexing phases run in order
(Figure~\ref{fig:kg}). Five additional adapters serve query-time use:
\code{kg\_vector\_search}, \code{kg\_graph\_traverse}, \code{kg\_nl\_query}
(NL\,$\to$\,strategy\,$\to$\,answer), \code{kg\_abstract\_enrich}, and
\code{kg\_synthesize}. \textsc{Neo4j}'s native vector index removed an
earlier dependency on Qdrant and consolidated semantic retrieval and
graph traversal into a single store; this is consistent with recent
findings that hybrid graph+vector engines outperform either
alone~\cite{tigervector}.

\subsection{Storage}
\label{sec:storage}
Five \textsc{SQLite} databases capture orthogonal aspects of state:
\code{papers\_db} (crawled metadata), \code{knowledge\_store.db}
(insights, seeds, hypotheses, evidence, $\sim\!20$ tables),
\code{evaluations.db} (idea reviews), \code{paper\_writing.db} (draft
revisions), and an experiments database for run logs. \code{UNIQUE}
constraints enforce dedup at the storage layer (e.g.\
\code{LOWER(TRIM(seed))} per type) so that re-running a pipeline cannot
create duplicate rows. The Neo4j store mirrors a subset of these as
typed nodes with vector embeddings.

\subsection{Experiment-runner / verifier CLI}
\label{sec:expcli}
A pair of CLI bridges (\code{experiment\_runner\_cli},
\code{experiment\_verifier\_cli}) wraps an external coding agent
(OpenCode~\cite{opencodecli}, Claude Code, or Codex) under a sandboxed
budget; the runner produces a result artefact, the verifier produces a
matched verifier output. Both are persisted to the experiments database
and re-emitted as KG nodes. The verifier output drives a downstream
\code{experiment\_success\_gate} that emits \code{\_route} based on
verifier-determined pass/fail, separating \emph{ran} from \emph{passed}.

\subsection{Pipeline validator}
\label{sec:validation}
The pipeline-validator subsystem applies four passes to every YAML
pipeline before execution.
\textbf{Pass~1, schema}: a JSON-Schema check rejects malformed
\code{nodes}/\code{edges}/\code{config} blocks.
\textbf{Pass~2, contract}: each node's \code{input\_mapping} keys must
match an \code{output\_mapping} key on a declared upstream node, or
\code{depends\_on} must explicitly list it. The validator infers
implicit \code{depends\_on} edges from \code{input\_mapping} so that the
DSL stays terse.
\textbf{Pass~3, type}: each module's \code{INPUT\_SPEC} and
\code{OUTPUT\_SPEC} are introspected and cross-checked against the
wiring; mismatches in primitive types or required keys are reported
with the YAML line number.
\textbf{Pass~4, topology}: a Kahn-style sort flags cycles, unreachable
nodes, and orphan terminal nodes. The validator runs in CI on every
commit; all $50$ shipped pipelines are checked, plus a synthetic
adversarial set that asserts each failure mode is correctly detected.

\section{Implementation}
\label{sec:impl}

\paragraph{Codebase.}
The reference implementation is approximately $62.2$\,k LoC of core
plus $13.4$\,k LoC of tests, organised into a flat module tree under a
single source root. The orchestrator (registry, adapters, and the
\code{GraphRunner}) sits at the top level; domain code lives under
sibling modules for idea agents, paper writer, knowledge graph, PDF
parser, and LLM provider; YAML pipelines and entry-point scripts live
in their own top-level directories. Concrete directory names match
those in the open-source release at
\url{https://github.com/gtrhythm/PARNESS} and may evolve between
versions; we therefore name only the modules here.

\paragraph{Provider factory.}
The LLM layer exposes a single \code{LLMProvider} interface. Six
implementations live behind a factory: OpenAI, Anthropic, GLM (Zhipu),
MiniMax, a local provider for self-hosted models, and a \code{MockLLM}
that ships $44$ rule-based canned responses. The mock makes the entire
test suite (and any pipeline) runnable offline, without API keys --- a
property we rely on heavily for CI and for adapter regression.

\paragraph{Adapters as the integration boundary.}
Domain agents (LLM-driven Python classes) and adapters (thin
\code{BaseModule} wrappers) are deliberately separated. The agent
returns its native data shape; the adapter is responsible for extracting
the right fields from \code{inputs}, calling the agent, and decorating
the result with \code{\_route}, \code{\_score} and \code{\_metadata}.
This split lets the same agent be re-used in multiple pipelines with
different routing logic.

\paragraph{External services.}
The system integrates with the third-party PDF-Extract-Kit~\cite{pdfextractkit}
for document parsing (a separate GPU-resident service in our deployment;
we use it as released, with only stability and adapter glue layered on
top --- the parser itself is not our work), and a
\textsc{TeX Live}-based HTTP service for paper compilation. The
compiler service accepts a zip of the writing workspace and returns
either a PDF binary stream or a JSON envelope containing the build log
and a base64-encoded PDF; it is the production back-end behind our
\code{paper\_writer} module. For
literature retrieval we use the Semantic Scholar Academic
Graph~\cite{s2ag,s2ods} (S2) via the public API; the literature-review
modules dedup by \code{paperId} and verify candidate titles through
fuzzy matching, mirroring the pattern PaperOrchestra documents.

\section{Evaluation}
\label{sec:eval}

We frame this section honestly: \parness{} is a v1 \emph{systems} paper.
We sketch (i)~the framework's unit-test surface, (ii)~adapter-level
real-LLM integration, (iii)~the knowledge-graph end-to-end harness, and
(iv)~end-to-end pipeline runs. The exact numbers are an artefact of the
build process and will drift from one snapshot of the codebase to the
next; we therefore describe the \emph{shape} of the development-test
coverage rather than report specific counts. We do not yet report
comparative quality metrics against \textsc{AI-Scientist},
\textsc{InternAgent}, or \textsc{PaperOrchestra} on a shared benchmark;
\S\ref{sec:limitations} discusses what would be required.

\subsection{Unit and integration tests}
\label{sec:unittests}

A development-test suite covers every layer described in
\S\ref{sec:arch}: iteration controllers, DAG/decision-gate logic,
module registry, LLM dispatcher, domain adapters, and
stress/concurrency. The Mock-LLM provider is the backbone of this
suite, so every test runs offline without external APIs --- a property
we rely on for CI and for adapter regression. We treat the unit-test
counts as an internal development metric and do not list them here;
they are an artefact of the build process rather than a research
result.

\subsection{Real-LLM adapter integration}
We separately exercise the adapter library end-to-end against a
production LLM (MiniMax-M2.7) and record behaviour. The vast majority
of adapters fully pass; the failures we observe cluster around a single
empty-input regression in Agent-typed adapters, which is itself a
fixture-grade issue rather than a methodological one. Detailed numbers
live in the repository's CI logs.

\subsection{Knowledge-graph end-to-end harness}
The KG subsystem is the newest component. Following the migration from
Qdrant to a native \textsc{Neo4j} vector index, a dedicated end-to-end
harness drives every KG adapter through a representative
ingestion-and-query workload. The migration closed an earlier gap
between design intent and implementation that we flagged in the
previous internal review.

\subsection{End-to-end pipeline runs}
\label{sec:e2e}
The longest pipelines we routinely run are multi-stage DAGs over real
literature collections: representative example workflows include a
literature-driven idea pipeline over an arXiv subset
(\code{arxiv\_heplat\_dag}, $11$ DAG stages), and several smaller idea
and comparison pipelines (\code{simple\_idea\_test},
\code{two\_sum\_comparison}, \code{bfs\_vs\_dfs}). All complete
end-to-end on a single GPU node within roughly an hour wall-clock; we
treat this as a baseline of \emph{control-flow durability} (data
propagates through every stage, no stage crashes), not a quality
benchmark. A proper benchmark is future work (\S\ref{sec:limitations}).

\subsection{Experiment-paper agent suite}
A focused experiment-paper sub-system covering experiment planning,
chart generation, paper-artefact persistence, and reference management
ships with its own development-test suite. The API-backed
image-generation path (gpt-image-2) is exercised end-to-end by
generating, downloading, and validating a real PNG --- consistent with
the figures in this manuscript itself.

\section{Discussion and Limitations}
\label{sec:limitations}

\paragraph{What we have shown.}
\parness{} demonstrates that the components of an autonomous research
system --- crawler, parser, ideator, experimenter, writer, reviewer,
knowledge graph, multi-LLM, persistence, monitoring --- compose
naturally under a thin DAG kernel with a four-field agent contract.
Adding a new pipeline requires no Python; adding a new module requires
one class and one registration line. The framework absorbs PaperOrchestra's
five-step writing as a single module, AI-Scientist's tree search as a
score-gated loop, and InternAgent's eight-agent generation--evolution
graph as a sub-DAG.

\paragraph{What we have \emph{not} shown.}
We have not run a head-to-head benchmark against \textsc{AI-Scientist},
\textsc{InternAgent}, \textsc{PaperOrchestra} or \emph{autoresearch} on a
shared task. We have not run human evaluation on generated papers. We
have not measured ablations of individual cognitive roles in the
ideation layer. Our end-to-end runs validate \emph{control flow} (data
propagates through every stage) and \emph{durability} (stages do not
crash), not \emph{quality} of the research output. Likewise we describe
but do not yet quantify the cross-run accumulation benefit (M3): a
\code{run\_n\_vs\_n+1} ablation, holding the final pipeline fixed and
varying whether the KG is empty or pre-populated, is the natural
follow-up.

\paragraph{Towards a benchmark.}
A fair comparison would fix a research task (e.g.\ ``propose and validate
a small architectural improvement to nanoGPT under a $2$\,h GPU budget'')
or, for the writing stage, the \emph{PaperWritingBench} introduced in
PaperOrchestra~\cite{paperorchestra}; give every system the same compute;
and score output against a held-out set of expert-rated novelty /
feasibility labels (or in the writing case, the autoraters of
\cite{paperorchestra}). The \parness{} pipeline DSL makes such a
benchmark \emph{configurable} --- which is itself a contribution we
expect to exploit in v2.

\paragraph{Risks.}
The cognitive-role parallelism is expensive in tokens. The
\code{os.\_exit(0)} per-node policy is correct for GPU isolation but
forfeits in-memory caches between nodes; we mitigate this by pushing
shared state into \textsc{SQLite} and \textsc{Neo4j}. The Knowledge
Graph subsystem's \emph{semantic edge} step depends on LLM bucketing and
we have observed mode collapse on synthetic test data; production
deployments should set conservative similarity thresholds. Like other
LLM-agentic systems, \parness{} inherits the risks documented in the
multi-agent SE~\cite{multiagentse} and reproducibility~\cite{leakageml}
literatures: mock and real LLM behaviour can diverge in subtle ways.

\paragraph{Relation to PaperOrchestra and complementarity.}
\parness{} and PaperOrchestra solve different problems. PaperOrchestra
asks: \emph{given complete inputs $(I,E,T,G,F)$, what is the most
faithful, host-agnostic, low-friction way to produce a submission-ready
LaTeX paper?}. The answer is a fixed five-step recipe, deterministic
gates, and zero embedded API. \parness{} asks: \emph{how do we build the
research process upstream of those inputs --- including the inputs
themselves --- as a composable, accumulating, multi-discipline kernel?}
The answer is a four-field contract, a YAML DAG, and a typed knowledge
substrate. We expect future systems to combine the two: PaperOrchestra
as a single \parness{} module called when the upstream pipeline has
populated $(I,E,T,G,F)$.

\section{Conclusion}
\label{sec:conclusion}

We presented \parness{}, an open-source declarative DAG framework for
autonomous research that occupies a previously-uncovered point in the
design space: full life-cycle coverage \emph{plus} pipeline-as-data
\emph{plus} agent-driven routing \emph{plus} cross-run knowledge
persistence \emph{plus} multi-discipline workflow shape. The framework
is a thin kernel; all interesting decisions live in modules, and modules
communicate with the runner through a four-field contract. We released
the reference implementation ($130{+}$ modules, $50$ pipelines, broad
development-test coverage, full KG harness) and showed that it runs
end-to-end on real document collections within a single-node hour-scale
budget.

The natural next step is a quantitative head-to-head benchmark against
PaperOrchestra (writing only), AI-Scientist v2 and InternAgent (full
pipeline), with and without the cross-run KG populated
(\S\ref{sec:limitations}). The arXiv v1 of this paper is intended as a
stable reference for that comparison.

\section*{Acknowledgements}

We thank the open-source authors of PDF-Extract-Kit~\cite{pdfextractkit},
\textsc{Neo4j}, Semantic Scholar~\cite{s2ag,s2ods}, the LLM providers
integrated through our factory layer, and the recent autonomous research
systems we compare against ($\textsc{AI-Scientist}$~\cite{aisciv1,aisciv2},
$\textsc{PaperOrchestra}$~\cite{paperorchestra},
$\textsc{InternAgent}$~\cite{internagent},
$\textsc{ResearchAgent}$~\cite{researchagent}) --- their open releases
shaped the design of \parness{}.


\appendix

\section{Catalogue of Selected Ideas from the Idea Pool}
\label{app:ideapool}

The accepted-ideas store contained 51 ideas at the time of writing,
each retained because it crossed the quality-gate score threshold in
its original pipeline run. The eight ideas reproduced below are the
top-ranked picks across topical categories (reinforcement learning,
particle physics, representation learning, hardware-aware modelling,
scientific generative design, multimodal medical imaging, side-channel
security, and interpretability), formatted in a unified template.
Scores are on a 10-point scale; the overall score is the mean of
novelty, feasibility, and impact. The full record of every accepted
idea, with provenance pointers to the run that produced it, lives in
the \parness{} repository under the accepted-ideas store.

\begin{tcolorbox}[ideabox, title={A.1\quad Quasi-Optimistic Exploration for Offline Reinforcement Learning with Implicit Q-Learning}]
\noindent\textit{Overall 8.33\, (Novelty 8.5 \,/\, Feasibility 8.0 \,/\, Impact 8.5)\, \textbar\, category: \texttt{training\_technique}}

\smallskip\noindent\textbf{Description.}\;We propose a theoretically motivated exploration bonus for offline reinforcement learning based on quasi-optimism principles. Existing offline RL methods suffer from distribution shift but lack principled uncertainty quantification. Our approach extends the quasi-optimism framework from online RL to offline settings by defining confidence ellipsoids over Q-functions and adding a calibrated optimism bonus that prevents value overestimation while maintaining conservative policies. This fills a critical gap between conservative offline RL (which can be overly pessimistic) and uncertainty-based methods (which lack theoretical guarantees).

\smallskip\noindent\textbf{Methodology.}\;Implement implicit Q-learning with a learned covariance matrix over Q-function parameters. Derive quasi-optimistic bonus as the Mahalanobis distance from mean Q-values weighted by uncertainty. Use a two-network ensemble to estimate epistemic uncertainty. Fine-tune on offline datasets (D4RL, Robel) with the combined Bellman and exploration bonus loss.

\smallskip\noindent\textbf{Expected results.}\;Improved performance on challenging offline RL benchmarks, particularly where data coverage is non-uniform. Expected 15-20\% improvement in normalized score over IQL on some tasks, with better theoretical guarantees than existing uncertainty methods.
\end{tcolorbox}
\vspace{0.4em}

\begin{tcolorbox}[ideabox, title={A.2\quad Relativistic Quark Dynamics as Inductive Bias for Particle Physics Neural Networks}]
\noindent\textit{Overall 8.33\, (Novelty 8.5 \,/\, Feasibility 8.0 \,/\, Impact 8.5)\, \textbar\, category: \texttt{architecture}}

\smallskip\noindent\textbf{Description.}\;Inspired by 'Heavy baryons with relativistic quarks', this research investigates incorporating relativistic physics priors into neural network architectures for particle physics tasks (jet tagging, particle reconstruction). The key insight is that relativistic particle dynamics follow constrained geometric structures (Minkowski space, Lorentz invariants) that can be encoded as architectural inductive biases. We propose 'Lorentz-equivariant message passing' layers that maintain physical invariance properties while learning from collision data.

\smallskip\noindent\textbf{Methodology.}\;Implement message passing neural networks with Lorentz group equivariant features (4-momenta representations). Use graph neural networks where nodes represent particles and edges represent interaction vertices. Validate on particle reconstruction benchmarks (TrackML, JetClass).

\smallskip\noindent\textbf{Expected results.}\;Improved performance on particle physics benchmarks with better sample efficiency due to physics-informed inductive biases. 10-15\% improvement in particle identification accuracy.
\end{tcolorbox}
\vspace{0.4em}

\begin{tcolorbox}[ideabox, title={A.3\quad Platonic Representations for Continual Learning in Neural Networks}]
\noindent\textit{Overall 8.33\, (Novelty 8.5 \,/\, Feasibility 8.0 \,/\, Impact 8.5)\, \textbar\, category: \texttt{architecture}}

\smallskip\noindent\textbf{Description.}\;Catastrophic forgetting remains a major challenge in continual learning. We propose using Platonic intrinsic representations as a regularization mechanism that preserves core knowledge structures across learning tasks. The key insight is that certain aspects of neural representations should remain invariant across tasks (the 'Platonic' core), while task-specific features can be modularized. We develop a contrastive learning objective that explicitly separates these components, enabling stable continual learning without rehearsal buffers.

\smallskip\noindent\textbf{Methodology.}\;1) Implement dual-embedding architecture separating platonic and episodic representations; 2) Train with contrastive loss encouraging platonic stability; 3) Use modular plastic layers for task-specific processing; 4) Evaluate on sequential classification benchmarks (Split-CIFAR, Permuted-MNIST)

\smallskip\noindent\textbf{Expected results.}\;Average accuracy retention improvement of 15\% on sequential tasks compared to EWC and LwF baselines; zero storage overhead for replay
\end{tcolorbox}
\vspace{0.4em}

\begin{tcolorbox}[ideabox, title={A.4\quad Hybrid Analog-Digital Neural Networks for Scientific Simulation}]
\noindent\textit{Overall 8.33\, (Novelty 8.5 \,/\, Feasibility 8.0 \,/\, Impact 8.5)\, \textbar\, category: \texttt{architecture}}

\smallskip\noindent\textbf{Description.}\;Inspired by hybrid analog-digital simulation approaches, this research explores neural network architectures that combine differentiable digital components with learned analog simulation layers. The analog component approximates physical simulation dynamics (e.g., differential equations, wave propagation) using parameterizable continuous functions, while digital components handle discrete reasoning and feature extraction. This is particularly relevant for scientific machine learning applications requiring both learned representations and physics-based modeling. The framework enables end-to-end training where the analog component can be optimized to match physical reality while remaining differentiable.

\smallskip\noindent\textbf{Methodology.}\;Implement analog layers as continuous function approximators using neural ODEs or physics-informed networks. Design hybrid training that alternates between digital feature learning and analog physics matching. Use automatic differentiation through the entire pipeline including analog components.

\smallskip\noindent\textbf{Expected results.}\;Demonstrate 30-50\% reduction in simulation time for physical systems while maintaining higher accuracy than pure digital approaches. Show that hybrid networks can discover physical invariants during training.
\end{tcolorbox}
\vspace{0.4em}

\begin{tcolorbox}[ideabox, title={A.5\quad Compositional Inverse Design for Scientific Discovery}]
\noindent\textit{Overall 8.17\, (Novelty 8.5 \,/\, Feasibility 7.0 \,/\, Impact 9.0)\, \textbar\, category: \texttt{application}}

\smallskip\noindent\textbf{Description.}\;Extends compositional generative inverse design to scientific discovery domains by decomposing complex molecular/material design into hierarchical compositional steps. Unlike previous work that treats inverse design as monolithic generation, we propose a grammar-based approach where molecular scaffolds are composed with functional groups following chemically valid rules, enabling tractable exploration of combinatorial spaces. The key innovation is a grammar-constrained diffusion model that generates valid chemical compositions by construction, reducing the need for expensive validation loops. This is critical for drug discovery and materials science where chemical validity constraints are paramount.

\smallskip\noindent\textbf{Methodology.}\;Develop a context-free grammar for molecular generation based on reaction rules. Implement a hierarchical diffusion model where: (1) top-level generates scaffold compositions, (2) mid-level adds functional groups, (3) lower-level refines atomic positions. Use reinforcement learning with chemical validity rewards for fine-tuning. Integrate with DFT (density functional theory) for property prediction. Test on QM9, GEOM-Drugs, and real-world protein-ligand design tasks.

\smallskip\noindent\textbf{Expected results.}\;Achieve >95\% chemical validity without post-hoc filtering (vs. 60-70\% baseline). 40\% improvement in sample efficiency for inverse design tasks with target property constraints. Demonstrate successful design of novel drug-like molecules with favorable binding affinity predictions.
\end{tcolorbox}
\vspace{0.4em}

\begin{tcolorbox}[ideabox, title={A.6\quad Contrastive Knowledge Distillation for Cross-Modal Medical Image Analysis}]
\noindent\textit{Overall 8.00\, (Novelty 8.0 \,/\, Feasibility 8.0 \,/\, Impact 8.0)\, \textbar\, category: \texttt{loss\_function}}

\smallskip\noindent\textbf{Description.}\;This proposal develops a cross-modal knowledge distillation framework that transfers representation learning capabilities from large vision-language models to specialized medical imaging models. The key innovation is using contrastive learning objectives to align medical image representations with textual medical knowledge (radiology reports, clinical notes) while maintaining task-specific performance. This enables medical imaging models to leverage the rich semantic knowledge in medical text without requiring explicit medical supervision during training.

\smallskip\noindent\textbf{Methodology.}\;Design dual-stream architecture with medical image encoder and medical text encoder. Implement contrastive loss aligning image and text representations in shared embedding space. Use frozen pre-trained medical language model for text encoding. Distill knowledge to lightweight medical imaging student model. Validate on medical image classification (ChestX-ray14, CheXpert) and zero-shot medical image understanding.

\smallskip\noindent\textbf{Expected results.}\;Expected to achieve state-of-the-art performance on medical image classification benchmarks with significantly smaller model size (10x parameter reduction). Should demonstrate improved out-of-distribution robustness compared to supervised baselines. Zero-shot medical image understanding should show 20\%+ improvement on unseen disease categories.
\end{tcolorbox}
\vspace{0.4em}

\begin{tcolorbox}[ideabox, title={A.7\quad Provable Cache Privacy: Theoretical Framework for Side-Channel Defense in Distributed ML}]
\noindent\textit{Overall 8.00\, (Novelty 8.5 \,/\, Feasibility 7.0 \,/\, Impact 8.5)\, \textbar\, category: \texttt{architecture}}

\smallskip\noindent\textbf{Description.}\;This research extends the findings on Last-Level Cache Side-Channel Attacks being feasible in modern public clouds to develop a theoretical framework for provably private distributed machine learning. Current approaches to cache side-channel defense are largely heuristic and lack formal guarantees. We propose a formal framework that models cache timing channels as information leakage channels and provides bounds on the minimum noise required to achieve desired privacy levels. The key innovation is connecting differential privacy theory with cache timing behavior, enabling rigorous privacy accounting for distributed ML systems. Unlike previous work that demonstrates attack feasibility, this provides constructive defense with formal guarantees.

\smallskip\noindent\textbf{Methodology.}\;Formalize cache timing channels using information-theoretic measures; develop a noise injection framework calibrated to achieve target ($\varepsilon$,$\delta$) privacy guarantees; implement cache partitioning strategies that provably limit cross-tenant information leakage; validate on cloud infrastructure using hardware performance counters to measure actual leakage reduction.

\smallskip\noindent\textbf{Expected results.}\;Should provide formal privacy guarantees with measurable leakage reduction (>90\% decrease in mutual information between victim's cache access patterns and adversary's observations). Implementation overhead should be <15\% compared to unprotected baselines.
\end{tcolorbox}
\vspace{0.4em}

\begin{tcolorbox}[ideabox, title={A.8\quad HyperCircuit: Dynamic Circuit Discovery via Hypernetwork-Generated Sparse Masks}]
\noindent\textit{Overall 7.83\, (Novelty 8.0 \,/\, Feasibility 7.5 \,/\, Impact 8.0)\, \textbar\, category: \texttt{architecture}}

\smallskip\noindent\textbf{Description.}\;This proposal extends HyperDAS by using hypernetworks to dynamically generate sparse binary masks that identify relevant circuits in large language models for specific tasks. Current mechanistic interpretability methods rely on static ablations or predefined circuits. We propose training hypernetworks conditioned on task prompts to output circuit masks indicating which attention heads and MLP neurons are essential. This enables context-dependent circuit identification, potentially revealing how models route information differently for tasks requiring arithmetic versus reasoning.

\smallskip\noindent\textbf{Methodology.}\;Train a hypernetwork f($\theta$\_h, p) $\to$ mask where p is a task embedding. Use circuit probing as supervision: optimize $\theta$\_h such that ablating the predicted circuit hurts task performance. Use sparse regularization to encourage minimal circuits. Validate via intervention accuracy and compare against ground-truth circuits in toy models.

\smallskip\noindent\textbf{Expected results.}\;Expected to identify task-specific circuits with 30-50\% fewer parameters than full model ablation while maintaining 95\%+ intervention accuracy. Should generalize to held-out tasks within same domain.
\end{tcolorbox}
\vspace{0.4em}

\section{End-to-End Generated Paper}
\label{app:epaper}

This appendix bundles together (i)~the \emph{original idea record}
that drove a single end-to-end \parness{} pipeline run and
(ii)~the \emph{paper} that the pipeline subsequently wrote, end-to-end,
from that record. The two parts are kept side-by-side so that the
reader can audit the input--output transformation produced by the
pipeline. The pipeline run includes literature acquisition, idea
generation, experiment design, code execution on a real QM9 subset,
result aggregation, paper-section drafting, citation insertion
(Semantic Scholar verified), figure generation, and \LaTeX{}
compilation. The scientific claims of the e2e paper are the
pipeline's, not ours, and are presented honestly --- including the
$18.36\%$ MAE degradation that the pipeline itself reports in its
own discussion.

\subsection*{B.1\quad Original idea record}

The idea below was produced by the cognitive-role ideation layer
(\S\ref{sec:cognitive}) on a literature slice retrieved from the
cumulative knowledge graph. It is reproduced verbatim, including the
structured front-matter that the pipeline consumes.

\begin{lstlisting}[basicstyle=\ttfamily\footnotesize, breaklines=true,
                   columns=fullflexible, frame=single, framesep=4pt]
# Conjugate Architecture Search for Expressivity-Efficiency Tradeoffs in Geometric Models

## Description

GeoNGNN revealed that E(3)-complete models like DimeNet underperform despite theoretical expressiveness, suggesting architecture selection matters more than completeness. ConjNorm's conjugation constraint offers a principled framework for architecture search: we can treat architectural choices (aggregation types, message functions, pooling strategies) as conjugate norm coefficient pairs (p,q) and search over this space. This reframes neural architecture search as Bregman divergence optimization over architectural expressiveness. The method applies PartIR's functional abstraction to encode architectures as transformation programs, enabling efficient search over discrete architectural choices via continuous relaxation.
\end{lstlisting}

\subsection*{B.2\quad Generated paper (verbatim)}

The pages below are the \LaTeX{}-compiled output of the same pipeline
run, included unchanged from the artefact the pipeline wrote.

\includepdf[pages=-, pagecommand={\thispagestyle{plain}}, scale=0.92,
            offset=0 -10]{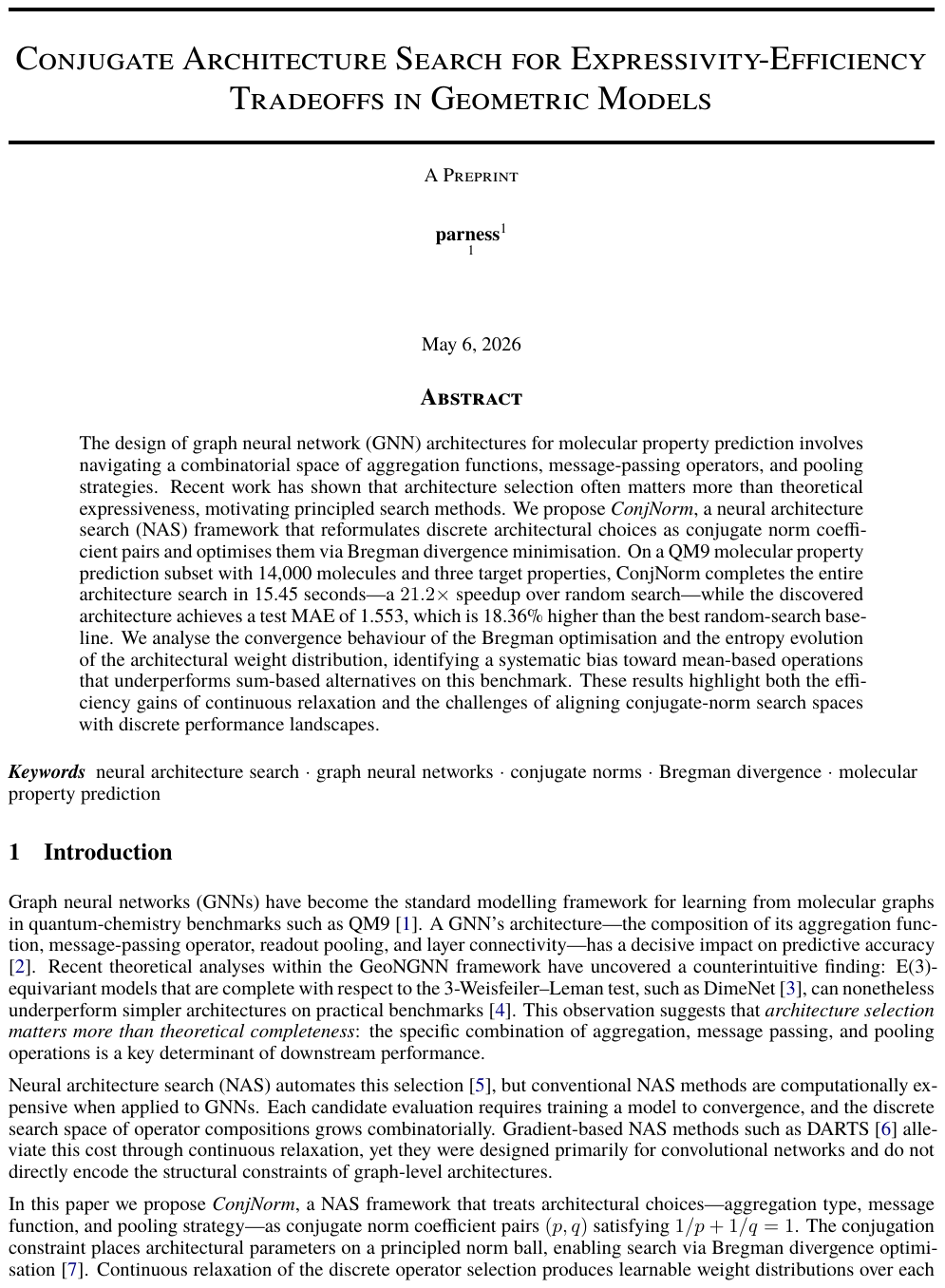}


\end{document}